\documentclass{aa}
\usepackage{txfonts}
\usepackage[dvips]{color}
\usepackage{graphicx}
\usepackage{natbib}
\bibpunct{(}{)}{;}{a}{}{,}
\usepackage{captcont}
\usepackage{subfigure}
\usepackage{hyperref}
\usepackage{sidecap}

\begin{document}

\title{Warm formaldehyde in the Oph IRS 48 transitional disk}
\titlerunning{Warm formaldehyde in Oph IRS 48}
\author{N. van der Marel\inst{1} 
\and E.F. van Dishoeck\inst{1,2}
\and S. Bruderer\inst{2}
\and T.A. van Kempen\inst{1}}
\institute{Leiden Observatory, Leiden University, P.O. Box 9513, 2300 RA Leiden, the Netherlands
\and Max-Planck-Institut f\"{u}r Extraterrestrische Physik, Giessenbachstrasse 1, 85748 Garching, Germany
}
\date{Accepted by A\&A, February 2014}
          
\abstract{Simple molecules such as H$_2$CO and CH$_3$OH in protoplanetary disks are the starting point for the
  production of more complex organic molecules. So far, the
  observed chemical complexity in disks has been limited because
of freeze-out of molecules onto grains in the bulk of the cold outer disk.}
  {Complex molecules can be studied more directly in transitional disks with large inner holes because these have a higher potential of
  detection through the UV heating of the outer disk and the directly exposed
  midplane at the wall.}{We used Atacama
  Large Millimeter/submillimeter Array (ALMA) Band 9 ($\sim$680 GHz) line data of the transitional
  disk Oph IRS 48, which was previously shown to have a large dust trap, to search for complex molecules in regions where planetesimals are forming.}{We report the detection
  of the H$_2$CO 9(1,8)--8(1,7) line at 674 GHz, which is spatially resolved as a semi-ring at $\sim60$ AU radius centered south from the star. The inferred H$_2$CO abundance is $\sim 10^{-8}$ , derived by
  combining a physical disk model of the source with a non-LTE excitation
  calculation. Upper limits for CH$_3$OH lines in the same disk give an
  abundance ratio H$_2$CO/CH$_3$OH $>$0.3, which indicates that
both ice formation and gas-phase routes play a role in the H$_2$CO production.
Upper limits on the abundances of H$^{13}$CO$^+$, CN and several other molecules in the disk were
also derived and found to be consistent with full chemical
models.}{The detection of the H$_2$CO line demonstrates the start of complex organic molecules in a planet-forming disk. Future ALMA observations are expected to reduce the abundance detection limits of other molecules by 1--2 orders of magnitude and test chemical models of organic molecules in (transitional) disks.} 


\maketitle

\section{Introduction}
Planets are formed in disks of dust and gas that surround
young stars. Although the chemical nature of the gas is simple, with
only small molecules such as H$_2$, CO, HCO$^+$ , or H$_2$CO detected so
far, the study of molecules in protoplanetary disks has resulted in a much better understanding of the origin of planetary
systems \citep[e.g.][]{WilliamsCieza2011,Henning2013}. Molecular line emission serves as a probe of disk properties,
such as density, temperature, and ionization. Furthermore, simple
species are the start of the growth of more complex organic and possibly prebiotic
molecules 
 \citep{Ehrenfreund2000,Mumma2011}.  Molecules in disks are incorporated
into icy planetesimals that eventually grow to comets and asteroids
that may have delivered water and organic material to
Earth. 
Therefore, a better understanding of the chemical composition of
protoplanetary disks where these icy bodies are formed provides
insight into the building blocks of comets and Earth-like planets
elsewhere in the Universe.

Protoplanetary disks have sizes of up to a few 100 AU, which makes them
similar to or larger than our own solar system ($\sim$50 AU
radius). However, at the distance of the nearest star-forming regions
these disks subtend less than a few arcsec on the sky, so that
telescopes with high angular resolution and high sensitivity are
needed to study their chemical composition. Most of the disks observed so far show 
no high chemical complexity \citep[e.g.][]{Dutrey1997,Thi2004,Kastner2008,Oberg2010}. In the surface layers of the disk, molecules are
destroyed by photodissociation by ultraviolet (UV) radiation from
the central protostar. In the outer disk and close to the disk
midplane, temperatures quickly drop to 100 K and lower, where all
detectable molecules, including CO, freeze out onto dust grains at
temperatures determined by their binding energies
 \citep{Bergin2007}. The chemical composition thus remains hidden in
ices. Only a very small part of the ice molecules is brought back into the gas
phase by nonthermal processes such as photodesorption. 

Transition disks have a hole in their dust distribution and thus
form a special
class of protoplanetary disks \citep{WilliamsCieza2011}. The hole allows a view into the usually hidden midplane composition
because the ices at the edge of this hole are directly UV irradiated by the
star, which results in increased photodesorption and thermal heating of the
ices \citep{Cleeves2011}. The hole is also an indicator
that the disk may be at the stage of forming planets. So far, the chemical composition of the outer regions of transitional disks appears to be similar to that of full disks, with detections of simple molecules including H$_2$CO \citep{Thi2004,Oberg2010,Oberg2011}. However, these data were taken with a typical resolution of $>$2''.

The study of the chemistry in protoplanetary disks has recently gained
much more perspective due to the impending completion of the Atacama
Large Millimeter/submillimeter Array (ALMA).  
ALMA allows us to study astrochemistry within protoplanetary disks at an
unprecedented level of complexity and at very small scales. It has the sensitivity to detect not only the dust, but also the gas inside dust gaps in transitional disks  \citep{vanderMarel2013,Casassus2013,Fukagawa2013,BruderervdM}.

The disk around Oph IRS 48 forms a unique laboratory for testing basic chemical processes
in planet-forming zones. IRS 48 is a massive young star ($M_* = 2
M_{\odot}$, $T_* \sim$10 000 K) in the Ophiuchus molecular cloud (distance 125 pc) with
a transition disk with a large inner dust hole, as revealed by mid-infrared imaging, which traces the hot small dust grains \citep{Geers2007}. The submillimeter continuum data (685 GHz or 0.43 mm)
from thermal emission from cold dust obtained with ALMA show that the millimeter-sized
dust is concentrated on one side of the disk, in contrast to the gas
and small dust grains. Gas was detected inside the dust gap down to 20 AU radius with ALMA data \citep{BruderervdM}, and 
strong PAH emission is also observed from within the cavity \citep{Geers2007}. The continuum asymmetry has been modeled as a major dust
trap \citep{vanderMarel2013}, triggered by the presence of a substellar
companion at $\sim$20 AU. The dust trap provides a region where dust
grains concentrate and grow rapidly to pebbles and then planetesimal sizes, producing
eventually what may be the analog of our Kuiper Belt. 

\citet{BruderervdM} presented and modeled the ALMA CO and continuum data,
together with complementary data at other wavelengths, to derive a
three-dimensional axisymmetric physical model of the IRS 48 disk. One
important conclusion is that the dust in the disk is warm, even out to
large radii, because the UV radiation can pass nearly unhindered
through the central hole. This implies that the dust temperature is
higher than 20 K throughout the disk so that CO does not freeze out
close to the midplane of the disk \citep{Collings2004,Bisschop2006}. 
The lack of a freeze-out zone of CO is important, since much of the
chemical complexity in a protoplanetary disks is thought to start with
the hydrogenation of the CO ice \citep{Herbst2009}. In the absence of
CO ice, gas-phase chemistry is the current main contributor to complex
molecule formation. Alternatively, the disk may have been colder in
the past before the hole was created, with the ices produced at that
time now evaporating. One of the simplest complex organic molecules,
H$_2$CO, can form both through hydrogenation of CO-ice
 \citep{Tielens1982,Hidaka2004,Fuchs2009,Cuppen2009} and through
gas-phase reactions.  H$_2$CO has been detected in several
astrophysical environments, such as the warm inner envelopes of low-
and high-mass protostars and protoplanetary disks including transitional disks
 \citep{Dutrey1997,Ceccarelli2000,Aikawa2003,Thi2004,Bisschop2007,Oberg2010,Oberg2011,Qi2013}. In
contrast with H$_2$CO, CH$_3$OH can only be formed through ice
chemistry \citep{Geppert2006,Garrod2006FD}, which means that the H$_2$CO/CH$_3$OH
ratio gives information on the H$_2$CO formation mechanism. Furthermore, H$_2$CO is a very interesting molecule for comparing the chemical
composition of disks, comets, and our solar system, because H$_2$CO- and CN-bearing molecules such as HCN and CN are precursors of amino acids. Synthesis of amino acids occurs in large 
asteroids in the presence of liquid water, for instance, through the Strecker synthesis route \citep{Ehrenfreund2000}.

In this work we present the detection of the H$_2$CO 9(1,8)--8(1,7)
line in IRS~48 down to scales of $\sim$30 AU, a high-excitation line originating from a level at 174
K above ground. We model its abundance and discuss the implications of
the origin of this molecule in combination with the nondetection of
other molecular lines. We were only able to detect this line thanks to the
tremendous increase in sensitivity at these high frequencies ($\sim$
670 GHz) using ALMA.

\section{Observations}

Oph IRS 48 ($\alpha_{\rm 2000}=$16$^{\rm h}$27$^{\rm m}$37\fs18, $\delta_{\rm 2000}=$-24\degr 30\arcmin 35.3\arcsec) was observed using the Atacama Large
Millimeter/submillimeter Array (ALMA) in Band 9 in the extended
configuration in Early Science Cycle 0. The observations were taken in
three observation execution blocks of 1.7 hours each in June and July
2012. During these executions, 18 to 21 antennas with baselines of up to
390 meters were used. The spectral setup contained four spectral
windows, centered on 674.00 , 678.84, 691.47, and 693.88 GHz. The
target lines of this setup were the $^{12}$CO $J$=6--5, C$^{17}$O
$J$=6--5, CN $N$=6$_{11/2}$--5$_{11/2}$, and H$^{13}$CO$^+$ $J$=8--7
transitions and the 690 GHz continuum. Each spectral window consists
of 3840 channels, a channel separation of 488 kHz, and thus a bandwidth
of 1875 MHz, which allows for serendipitous detection of other
lines. The resulting velocity resolution is 0.21 km s$^{-1}$ (for a
reference of 690 GHz).  Table \ref{tbl:targetlines} summarizes the
observed lines and frequencies.  For CH$_3$OH, another transition was covered in our spectral setup at 678 GHz (4(2,3)--3(1,2)) but with the same Einstein coefficient and $E_U$ as the 674 GHz transition. For CN, the
6$_{11/2}$--5$_{11/2}$ component covered in our observations has a rather low Einstein A coefficient (averaged over the unresolved
hyperfine components). The CN lines at 680 GHz with Einstein A values stronger by two orders of
magnitude were just outside the correlator
setting, which was optimized for $^{12}$CO and C$^{17}$O 6--5.

We reduced and calibrated the data using the Common Astronomy Software Application (CASA) v3.4. The bandpass was calibrated using quasar 3c279, and fluxes were calibrated against Titan. For Titan, the flux was calibrated with a model that was fit to the shortest one-third of the baselines, since Titan was resolved out at longer baselines and the fit to the model would not be improved. The absolute flux calibration uncertainty in ALMA Band 9 is $\sim$20\%. 

The resulting images have a synthesized beam of 0.32'' $\times$
0.21'' or 38$\times$25 AU (1.87$\cdot10^{-12}$ sr) and a position angle of
96$^{\circ}$ after applying natural weighting and cleaning. After extracting the continuum data and the line
data of $^{12}$CO $J$=6--5 and C$^{17}$O $J$=6--5, we performed a
search for lines of other simple and complex molecules within our
spectral setup, but apart from H$_2$CO, no other convincing features
were found. A rectangular cleaning mask centered on the detected
emission peaks was used during the cleaning process of the $^{12}$CO line and was adapted to the detected emission of each channel. The
final channel-dependent mask was then applied to the frequencies of
the targeted weaker lines, given in Table \ref{tbl:targetlines}. The
final rms level was 20-30 mJy beam$^{-1}$ per 1 km s$^{-1}$ channel,
depending on the exact frequency.

For more details on the data reduction see the supplementary online material of \citet{vanderMarel2013}.

\begin{table}[h]
\small
  \caption{Molecular lines in addition to CO observed toward IRS48 (parameters taken from the Cologne Database for Molecular Spectroscopy \citep{Muller2001,Muller2005})}
  \label{tbl:targetlines}
  \begin{tabular*}{0.5\textwidth}{lllrl}
    \hline
    Molecule & Transition & Rest frequency & $E_u$ & $^{10}$log($A_{u\ell}$)\\
    &&(GHz)&(K)&\\
    \hline
        H$_2$CO&9(1,8)--8(1,7) & 674.80978 & 174 & -2.04\\
        CH$_3$OH&4(2,2)--3(1,3)&678.78546&61&-2.87\\
        H$^{13}$CO$^+$&8--7&693.87633&150&-1.56\\
        CN&6$_{11/2}$--5$_{11/2}$&678.84488&114&-5.35\\
        $^{34}$SO$_2$&13(5,9)--12(4,8)&691.99178&143&-2.67\\
        C$^{34}$S&14--13&674.47362&180&-2.27\\
        HNCO&9(1,9)--10(0,10)&678.23827&91&-2.80\\
        c-C$_3$H$_2$&10(10,1)--9(9,0)&693.68725&182&-1.77\\
        N$_2$D$^+$&9--8&693.80616&167&-1.42\\
        D$_2$O&4(2,2)--4(1,3)&692.24358&236&-1.80\\     
    \hline
  \end{tabular*}
\end{table}

\section{Results}
\begin{figure}[!ht]
\includegraphics[scale=0.4]{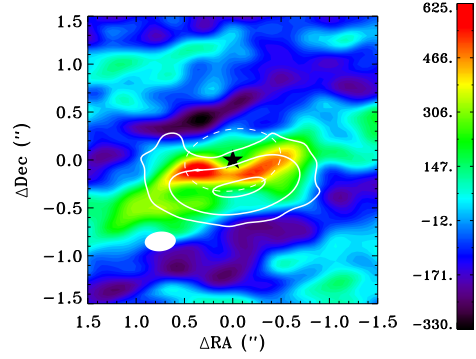}
\caption{Integrated intensity map of the H$_2$CO 9(1,8)--8(1,7)
  emission. The color bar gives the flux scale in mJy beam$^{-1}$ km s$^{-1}$. 
  The white contours indicate the 690 GHz continuum at 3, 30, and
  300 $\sigma$ from the thermal emission of the millimeter-sized dust
  grains. The position of the star IRS 48 is indicated by the star and
  the 60 AU radius (the dust trap radius) by a white dashed ellipse.}
\label{fig:momentmap}
\end{figure}

\begin{figure}[!ht]
\includegraphics[scale=0.4]{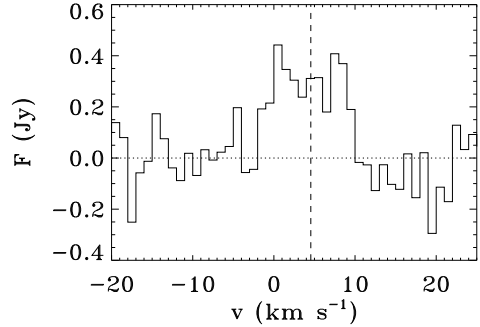}
\caption{Spectrum of H$_2$CO 9(1,8)-8(1,7) toward IRS 48 
spatially integrated over the
  rectangle [-1.0'' to +1.0'', -0.3'' to +0.2'']. The dotted line indicates the zero-emission line and the dashed line the source velocity of 4.55 km s$^{-1}$. 
}
\label{fig:spectrum}
\end{figure}

\begin{figure*}[!ht]
\includegraphics[scale=0.9]{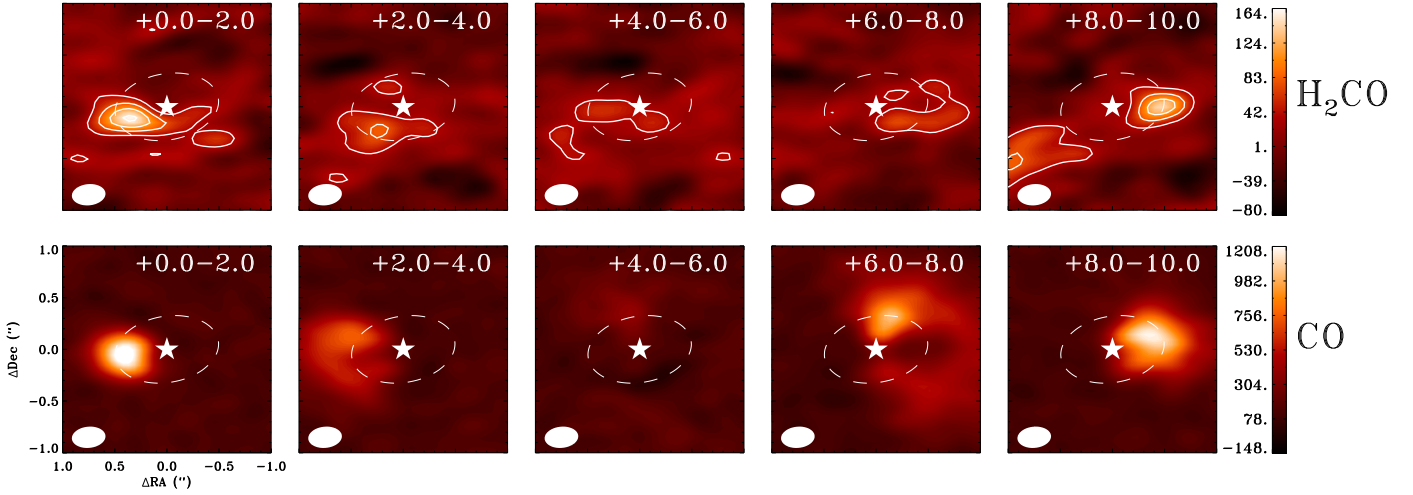}
\caption{Comparison of channel maps H$_2$CO 9(1,8)--8(1,7) (top row)
  with $^{12}$CO 6--5 (bottom row) (data taken from \citet{BruderervdM}. The color bar indicates the fluxes
  in mJy beam$^{-1}$. The number in the upper right corner of each map is the
  velocity range of that map. In the H$_2$CO maps, the white contours
  indicate the 20\%, 40\%, 60\%, 80\%, and 100\% of the peak intensity
  of all channels (the peak intensity is 195 mJy beam$^{-1}$). The dashed ellipse indicates the 60 AU radius
  ring. The CO +4--+6 km s$^{-1}$ channels are affected by foreground absorption.}
\label{fig:channelmap}
\end{figure*}

The H$_2$CO 9(1,8)--8(1,7) line was detected at 3--8$\sigma$ levels in
the channels between $v_{\rm LSR}$ = -0.5 and +9.5 km s$^{-1}$ \citep[the source
velocity is 4.55 km s$^{-1}$,][]{vanderMarel2013}. The integrated
flux between -0.5 and +9.5 km s$^{-1}$ shows a flattened structure
centered just south of the star (Figure \ref{fig:momentmap}). The stellar
position (Table 1) was determined from the fastest velocity channels
where the $^{12}$CO emission from the same data set was detected. The emission
is extended across the spatial region, corresponding to a rectangle
[-1.0'' to +1.0'', -0.3'' to +0.2''] having an area of
2.1$\cdot10^{-11}$ sr. Across this area, the integrated flux is found to
be $\sim$3.1 $\pm$ 0.6 Jy km s$^{-1}$ (see Figure \ref{fig:momentmap} and \ref{fig:spectrum}). The emission is, where detected, cospatial with
the $^{12}$CO 6--5 emission at the same velocity channels, although the northern half of the emission is missing (see Figure
\ref{fig:channelmap}). Gas orbiting a star in principle follows
Kepler's laws, with the velocity $v$ depending on the orbital radius
$r$ according to $v = \sqrt{{GM}/{r}}$, with $G$ the gravitational
constant and $M$ the stellar mass. The CO 6--5 emission is found to follow Keplerian motion in an
axisymmetric disk \citep{BruderervdM}. The emission in the high-velocity channels is significantly stronger than the velocities closer
to the source velocity, which can be partly explained by limb
brightening due to the 50$^{\circ}$ inclination of the disk.  Some H$_2$CO and CO emission is found in various
channels in the southeast corner (at velocities of +1--2 km s$^{-1}$)
 and not follow the Keplerian motion pattern \citep{BruderervdM},
but because of the low $S/N$ it is not possible to distinguish whether
this emission is real or a cleaning artifact for the case of H$_2$CO. 

A striking aspect of these images is that H$_2$CO is only detected south of the star,
as is also found for the submillimeter continuum, which shows a north-south
asymmetry of the disk with a contrast factor of $>$100. The
north-south contrast of H$_2$CO can only be constrained to be a factor
$>$2 because of the low $S/N$ of this detection. Moreover, the H$_2$CO does
not follow the submillimeter continuum exactly: at the continuum peak no H$_2$CO emission is detected at all. The velocity channels in
$^{12}$CO between 2.5 and 4.5 km s$^{-1}$ suffer from absorption by
foreground clouds, but the H$_{2}$CO abundance and excitation in these
clouds is too low to absorb the H$_{2}$CO disk emission in this highly
excited line. Thus, the absence of H$_2$CO at the peak millimeter
continuum is significant.

The other targeted molecules remain undetected, but the non-detection
provides 3$\sigma$ upper limits that can be used in the
models. For the 25 mJy beam$^{-1}$ km s$^{-1}$ rms level, we
estimated the upper limit on the total flux of CH$_3$OH as follows: for the H$_2$CO detected emission, the total
surface area d$\Omega$ with emission $>3\sigma$ is $\sim
1.3\cdot10^{-11}$ sr or $\sim$7 beams, and the detectable emission lies
between -0.5 and +9.5 km s$^{-1}$ (11 channels). The 3$\sigma$ upper
limit on the total flux is thus
\begin{equation}
F_{\rm upp} = \sqrt{d\Omega/\Omega_{\rm beam}} \cdot 3\sigma_{\rm rms} \cdot 1.2\sqrt{11\cdot1} = 0.79 \ {\rm Jy \ km \ s^{-1}}
,\end{equation}
where the factor 1.2 is introduced to compensate for small-scale noise variations and calibration uncertainties at this high frequency of the observations. Note that this is a conservative limit because the H$_2$CO emission is most likely narrower in the radial direction than the spatial resolution. 
For the CN, H$^{13}$CO$^+$ , and other molecules in Table \ref{tbl:targetlines} we expect the emission to be cospatial with the CO emission, which covers $\sim4.8\cdot10^{-11}$ sr or $\sim26$ beams and integrated between -3 and +12 km s$^{-1}$. The upper limit is thus
\begin{equation}
F_{\rm upp} = \sqrt{d\Omega/\Omega_{\rm beam}} \cdot 3\sigma_{\rm rms} \cdot 1.2\sqrt{15\cdot1} = 1.8 \ {\rm Jy \ km \ s^{-1}}
.\end{equation}

\section{Model}
\subsection{Physical structure}

Analysis of the abundance and spatial distribution of the H$_2$CO in
the disk requires a physical model of the temperature and gas density
as a function of radius and height in the disk. We used the best-fitting model of the gas structure from \citet{BruderervdM} based on
the $^{12}$CO and C$^{17}$O 6--5 lines from the same ALMA data set and
the dust continuum. The proper interpretation of the gas
disk seen in $^{12}$CO 6--5 emission requires a thermo-chemical disk
model, in which the heating--cooling balance of the gas and chemistry
are solved simultaneously to determine the gas temperature and
molecular abundances at each position in the disk. Moreover, even though
the densities in disks are high, the excitation of the rotational
levels may not be in thermodynamic equilibrium, and there are steep
temperature gradients in both radial and vertical directions in the
disk. The DALI model \citep{Bruderer2012,Bruderer2013} uses a combination of a stellar photosphere with a disk density distribution as input. For  IRS 48, the stellar photosphere is represented by a blackbody of 10 000 K. It solves for the dust temperatures
through continuum radiative transfer from UV to millimeter
wavelengths and calculates the chemical abundances, the molecular
excitation, and the thermal balance of the gas. It was developed for
the analysis of the gas emission structures such as are found for
transitional disks \citep{Bruderer2013}.

DALI uses a reaction network described in detail in
\citet{Bruderer2012} and \citet{Bruderer2013}. It is based on a
subset of the UMIST 2006 gas-phase network
\citep{Woodall2007}. About 110 species and 1500 reactions are
included. In addition to the gas-phase reactions, some basic
grain-surface reactions (freeze-out, thermal and nonthermal
evaporation and hydrogenation like g:O $\to$ g:OH $\to$ g:H$_2$O and
H$_2$/CH$^+$ formation on PAHs) are included. The g:X notation refers to atoms and molecules on the grain surface. The photodissociation
rates are obtained from the wavelength-dependent cross-sections by
\citet{vanDishoeck2006}. The adopted cosmic-ray ionization rate is
$\zeta=5\times 10^{-17}$ s$^{-1}$. X-ray ionization and the effect of
vibrationally exited H$_2$ are also included in the network.  The
model chemistry output will be used to compare with the HCO$^+$ and CN
data. However, since no extensive grain-surface chemistry is included,
it is not suitable to model the H$_2$CO and CH$_3$OH chemistry.

\begin{figure}[!ht]
\begin{center}
\includegraphics[scale=0.7]{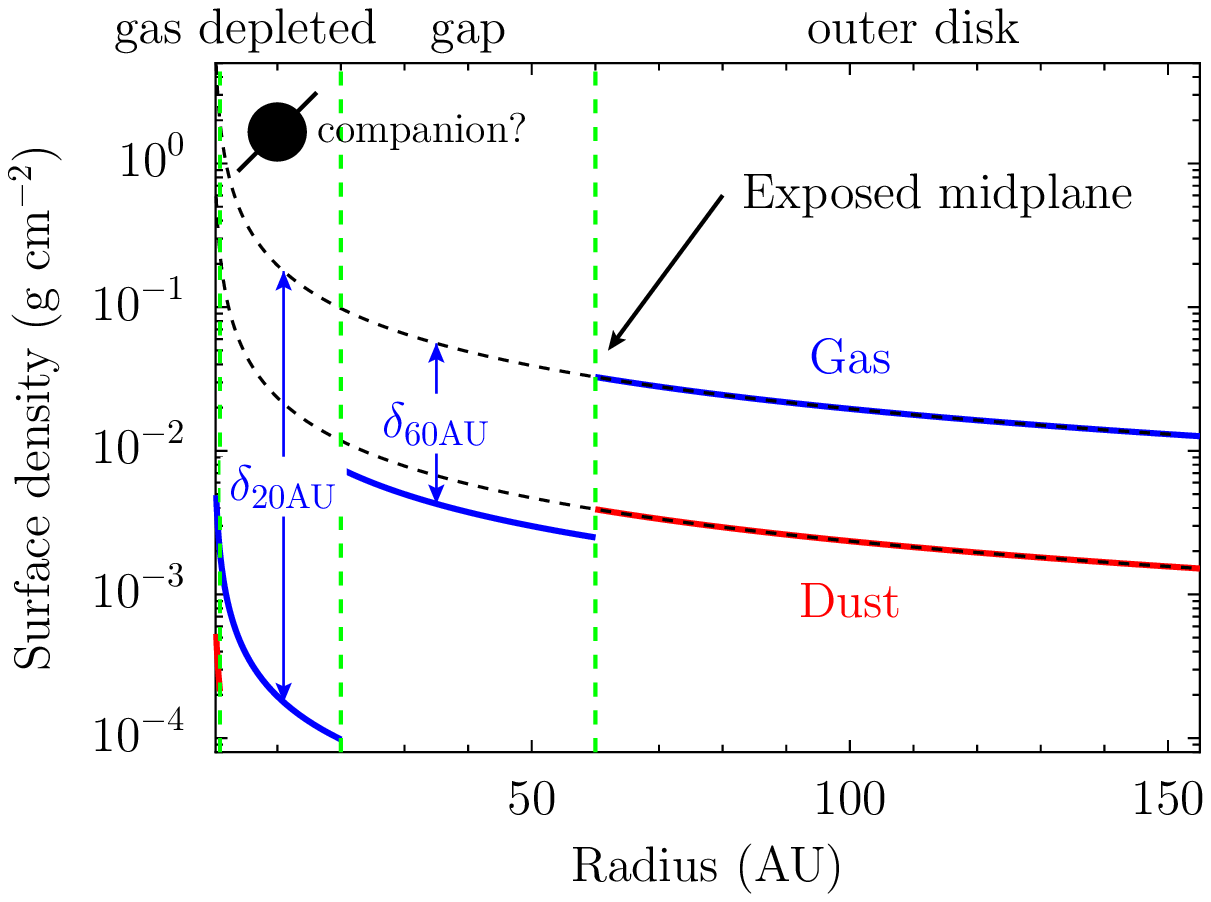}
\caption{Adopted radial density structure of the IRS 48 disk, taken from
  \citet{Bruderer2013}. The blue lines indicate the gas surface density, the red lines the dust surface density. The dotted black lines show the undisturbed surface density profiles if it continued from outside inwards without depletion. The green dashed lines indicate the radii at which the depletions start.}
\label{fig:disksetup}
\end{center}
\end{figure}

The surface density profile for the best-fitting model for IRS 48 \citep{BruderervdM} 
including the gap is shown in Figure \ref{fig:disksetup}.  It is found
that the gas disk around IRS~48 has a very low mass (1.4$\cdot10^{-4}$
M$_{\odot}$ = 0.15 M$_{\rm Jup}$) compared with the mean disk mass of
$\sim$5 M$_{\rm Jup}$ for normal disks
 \citep{WilliamsCieza2011,Andrews2013}. Furthermore, the disk has a large scale-height, allowing a large portion of the inner walls to be irradiated by the star.
The radial gas structure has two density drops: a drop
$\delta_{20}$ of $<10^{-1}$ in the inner 20 AU (probably caused by a planetary or substellar companion) and an additional
drop $\delta_{60}$ of 10$^{-1}$ at 60 AU radius.

\begin{figure*}[!ht]
\begin{center}
\subfigure{\includegraphics[scale=0.4,trim=0 40 40 0]{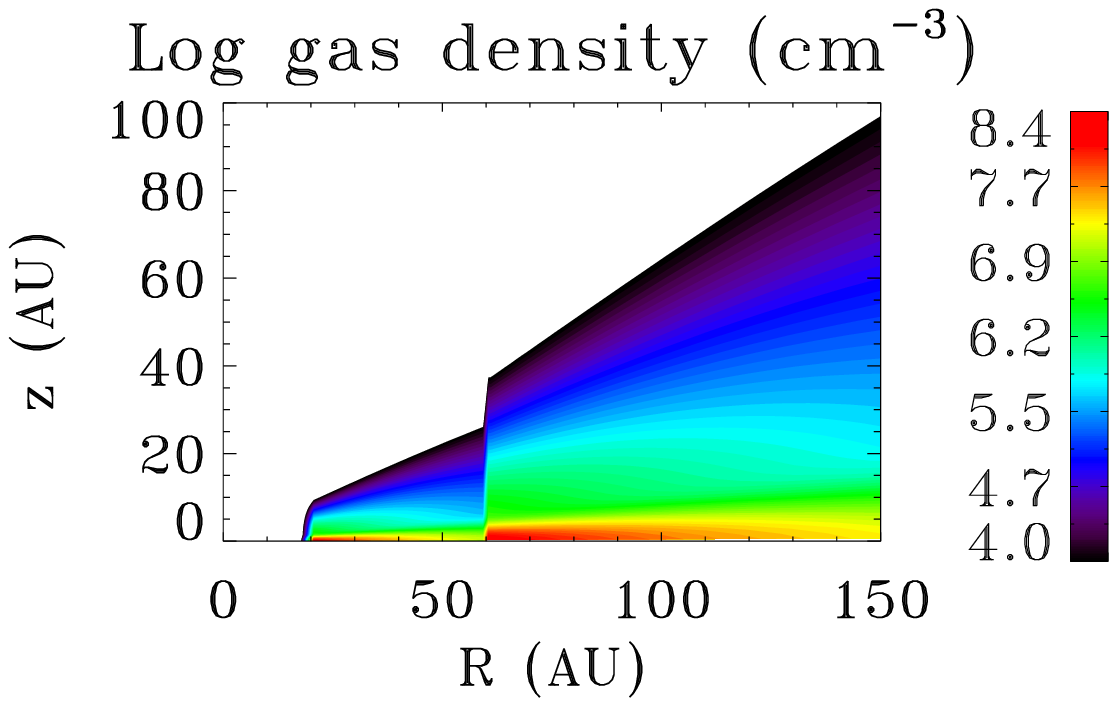}}
\subfigure{\includegraphics[scale=0.4,trim=0 40 40 0]{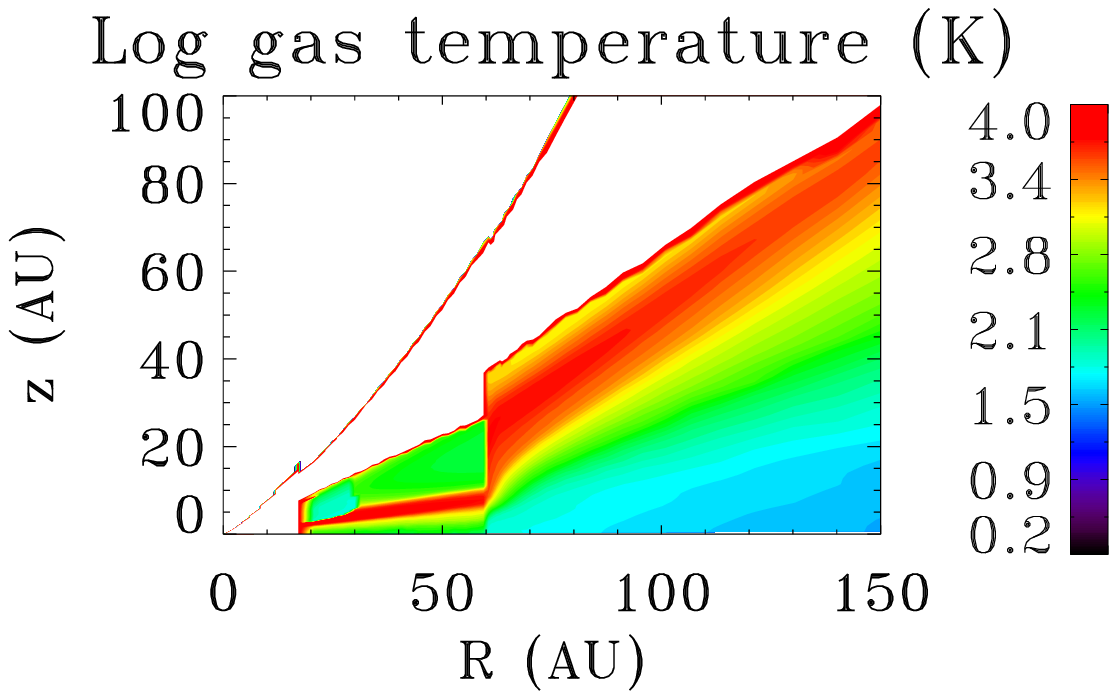}}
\subfigure{\includegraphics[scale=0.4,trim=0 40 40 0]{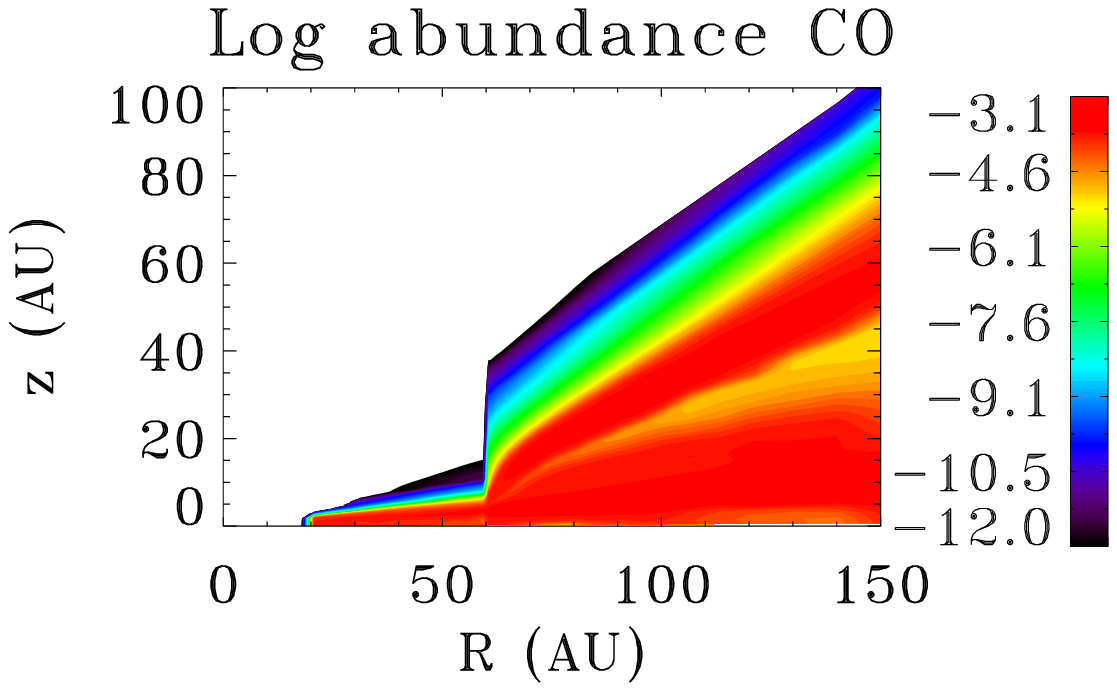}}\\
\subfigure{\includegraphics[scale=0.4,trim=0 40 40 0]{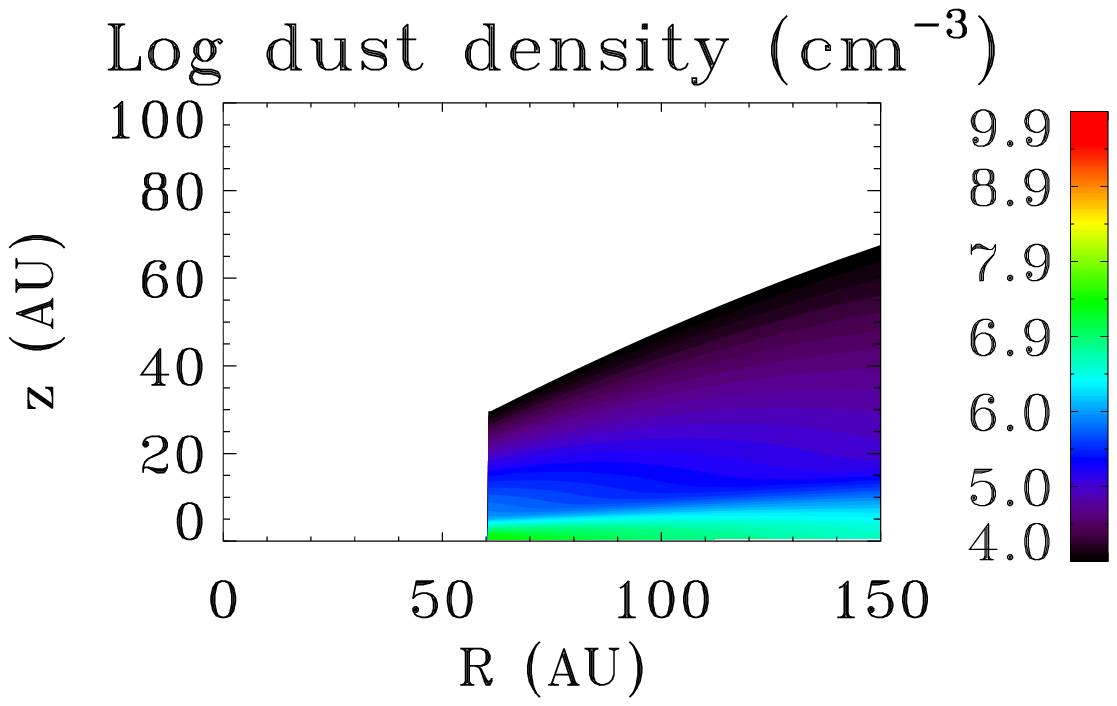}}
\subfigure{\includegraphics[scale=0.4,trim=0 40 40 0]{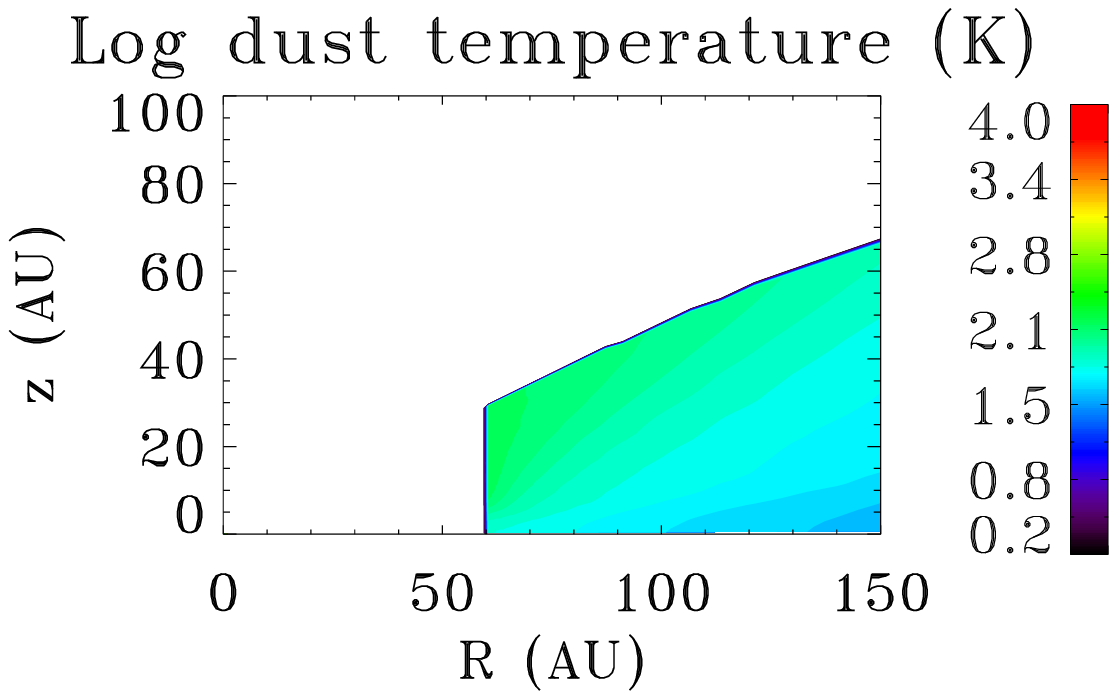}}
\subfigure{\includegraphics[scale=0.4,trim=0 40 40 0]{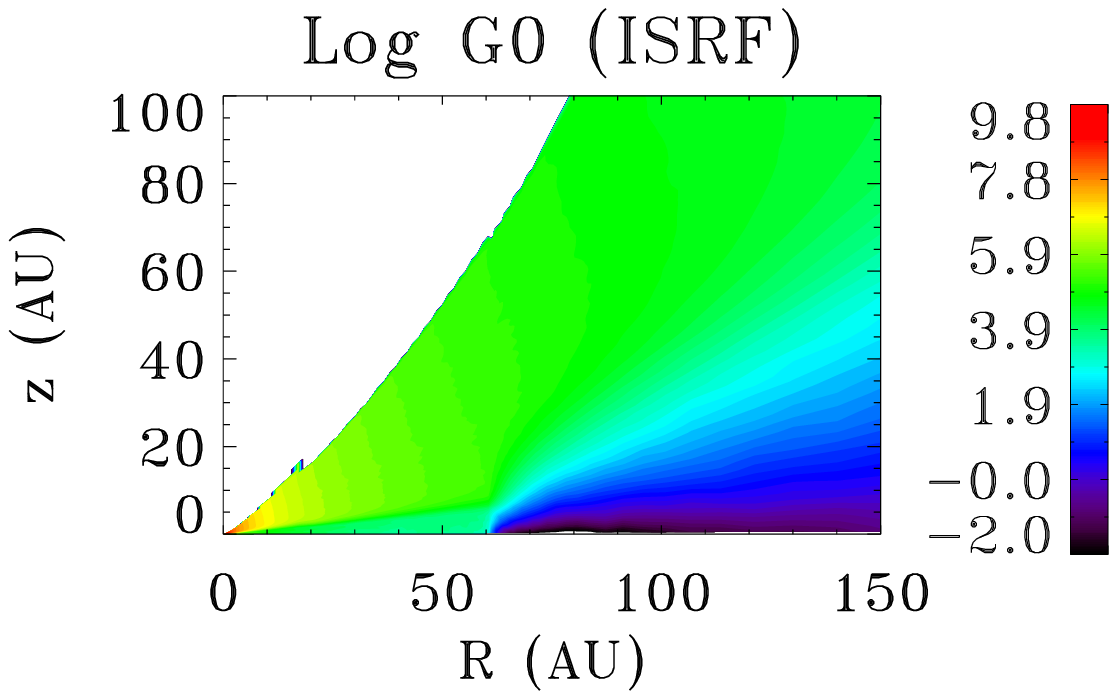}}
\end{center}
\caption{Physical-chemical model of IRS 48 from \citet{BruderervdM}. The panels show gas density (cm$^{-3}$), gas temperature (K), CO abundance with respect to H$_2$, dust density (cm$^{-3}$), dust temperature (K), and UV field in $G_0$ ($G_0$ = 1 refers to the standard interstellar radiation field). The given numbers are in $^{10}$log scale.}
\label{fig:modelpars}
\end{figure*}

The resulting densities, temperatures and CO abundances of the model
are given in Figure \ref{fig:modelpars}. The gas density is 10$^{5-6}$
molecules cm$^{-3}$ in the upper layers of the disk, increasing to
10$^8$ cm$^{-3}$ close to the midplane near 60 AU.  The gas
temperature is typically a few 100 K, except in the upper layers where
the temperature reaches several thousand K. The UV radiation field is
enhanced by factors of up to 10$^8$ in the disk, indicated by
$G_0$ (panel 6 in Figure \ref{fig:modelpars}). $G_0$ = 1 refers to the
interstellar radiation field defined as in \citet{Draine1978},
$\sim$2.7 $\times$ 10$^{-3}$ erg s$^{-1}$ cm$^{-2}$ with
photon-energies in the far UV range between 6 eV and 13.6 eV. The CO
abundance is $\sim10^{-4}$ compared with H$_2$ throughout the bulk of the outer
disk and is not frozen out.

\begin{figure*}[!ht]
\hspace{4cm}Model 1\hspace{3.5cm}Model 2\hspace{5cm}Model 3
\begin{center}
\subfigure{\includegraphics[scale=0.45,trim=0 40 0 0]{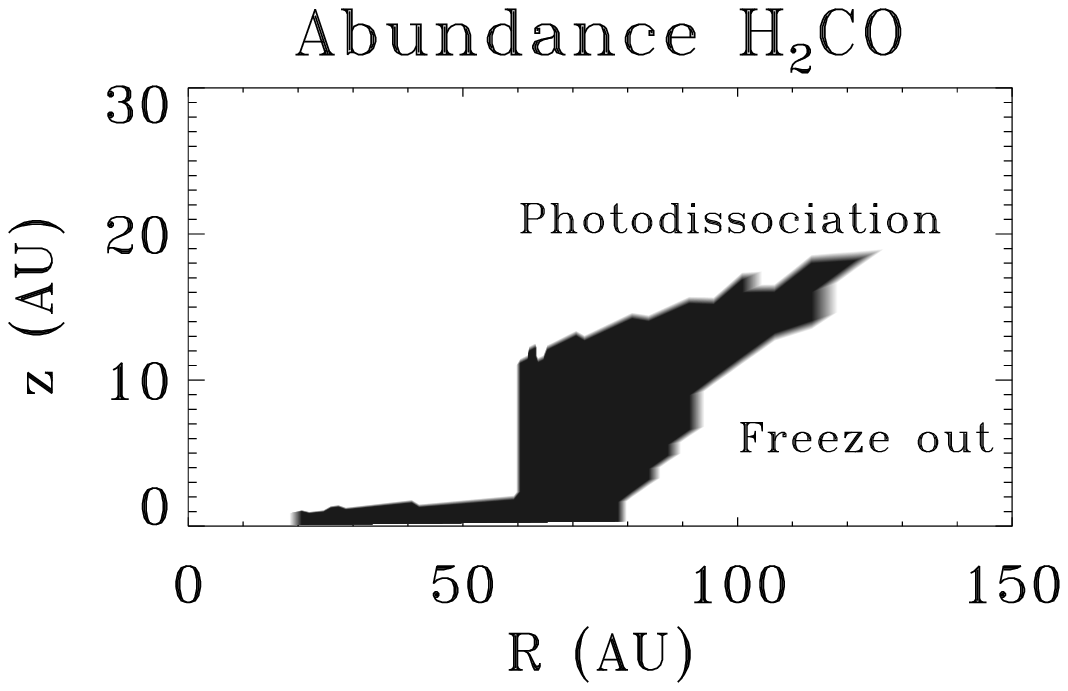}}
\subfigure{\includegraphics[scale=0.45,trim=0 40 40 0]{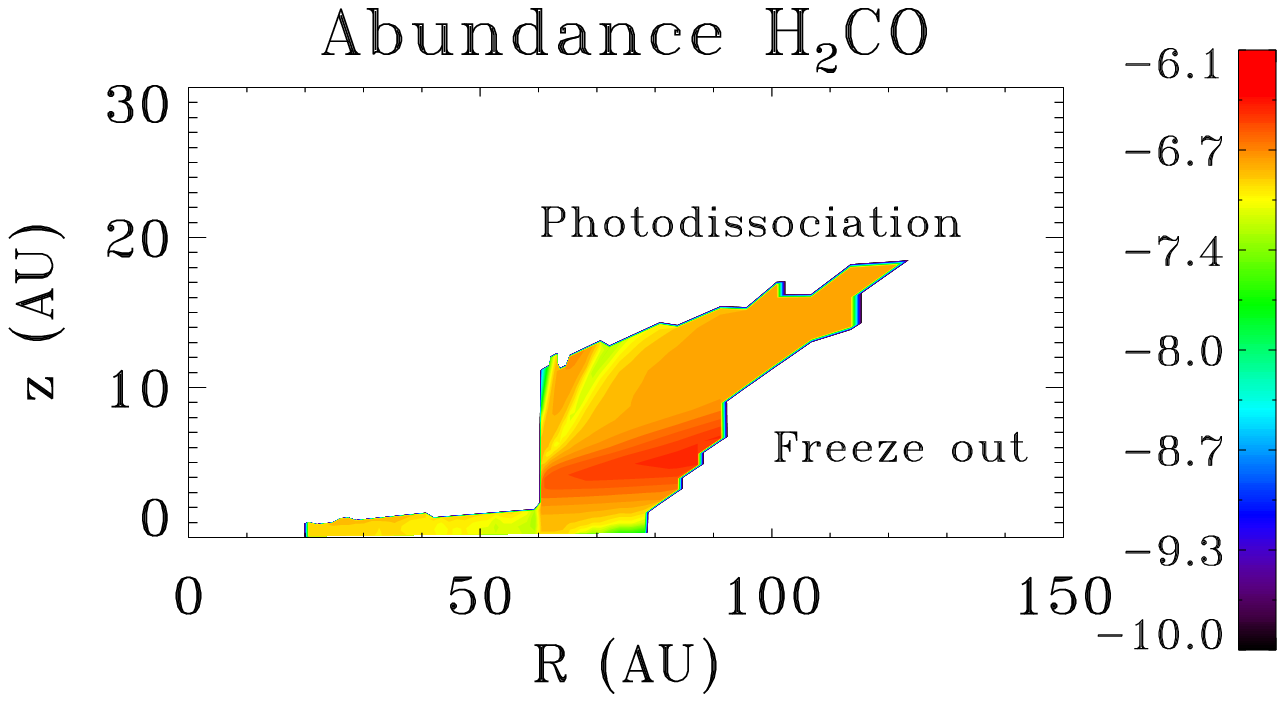}}
\subfigure{\includegraphics[scale=0.45,trim=0 40 40 0]{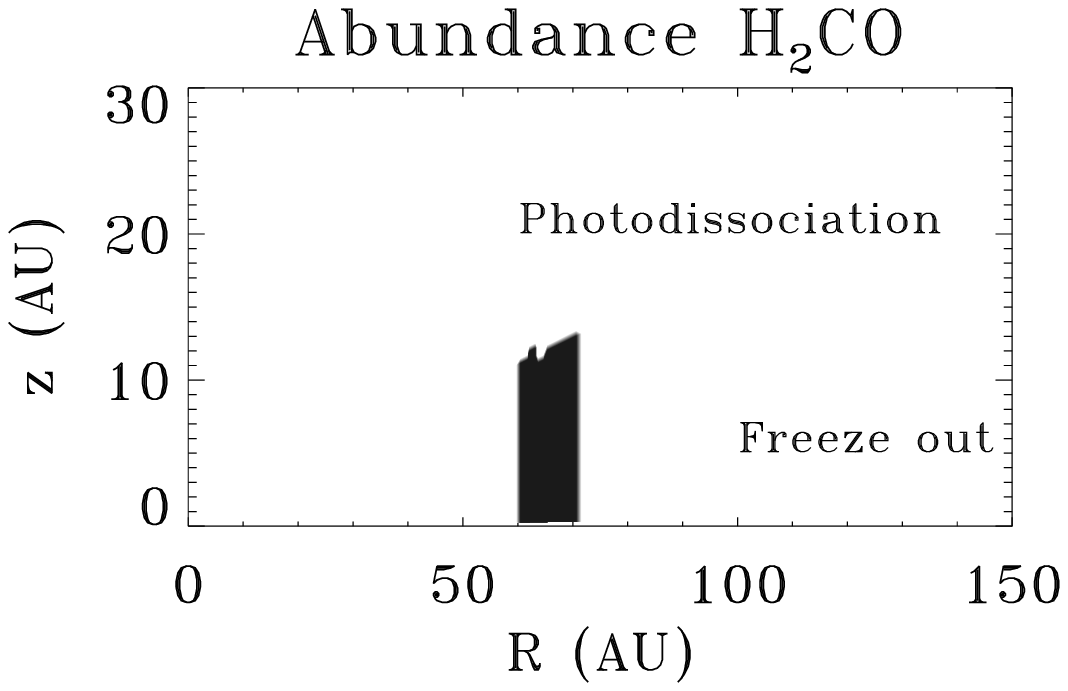}}
\caption{Trial abundance models 1, 2 and 3 for H$_2$CO. The H$_2$CO abundance is limited by photodissociation in the upper layer and freeze out below 60 K. Model 1 assumes a constant abundance, model 2 assumes a fractional abundance with respect to CO, model 3 assumes a constant abundance in between 60 and 70 AU radius and zero abundance at other radii.}
\label{fig:abunprofiles}
\end{center}
\end{figure*}

The H$_2$CO emission was modeled using the LIne Modeling Engine
(LIME), a non-LTE excitation spectral line radiation transfer code
 \citep{Brinch2010}. The physical structure described above is used as
input. The first step in the analysis is to empirically constrain the H$_2$CO
abundance by using three different trial abundance profiles guided by
astrochemical considerations.  The inferred abundances were then {\it a
  posteriori} compared with those found in the full chemical
models. In model 1, the abundance was assumed to be constant throughout
the disk, testing abundances between 10$^{-5}$ and
10$^{-11}$ with respect to H$_2$. In model 2, the abundance was taken
to follow the CO abundance calculated by the DALI model, taking a
fractional abundance ranging between 10$^{-3}$ and 10$^{-8}$ with
respect to CO.  Model 3 was inspired by \citet{Cleeves2011} by
setting the H$_2$CO to zero except for a ring between 60 and 70
AU. This model assumes that the UV irradiated inner rim has an
increased
chemical complexity that can be observed directly, as material from
the midplane has been liberated from the ices. The abundance profile is additionally constrained by the photodissociation and freeze out. H$_2$CO can only exist below
the photodissociation height, which was taken as the height ($z$ direction) where a
hydrogen column density of $N$(H$_2$) = $4\cdot10^{20}$ cm$^{-2}$ is
reached. At this column, the CO photodissociation rate drops
significantly due to shielding by dust as well as self-shielding and
mutual shielding by H$_2$ for low gas-to-dust ratios
 \citep{Visser2009}. Therefore, at each radius this value was calculated
and the abundance was set to zero above it. For radii $<$60 AU, H$_2$CO
is photodissociated almost entirely down to the midplane because of
the lower total column density. Furthermore, H$_2$CO is expected to
be frozen out on the grains at temperatures $<$60 K since it has a
higher binding energy \citep{Ioppolo2011} than CO, therefore the abundance in regions below this temperature were also set to
zero. All three abundance profiles are
shown in Figure \ref{fig:abunprofiles}.

The LIME grid was built using linear sampling, with the highest grid
density starting at 60 AU, using 30 000 grid points and 12 000 surface
grid points, using an outer radius of 200 AU. The image cubes were
calculated for 60 velocity channels of width 0.5 km s$^{-1}$ spectral
resolution, in 5''$\times$5'' maps with 0.025'' pixels.
Collisional rate coefficients were taken from the Leiden Atomic and Molecular Database (LAMDA) \citep{Schoier2005} with references to the original collisional rate coefficients as follows: H$_2$CO \citep{Tros2009}, CH$_3$OH \citep{Rabli2010}, H$^{13}$CO$^+$ \citep{Flower1999}, and CN \citep{Lique2010}. For the other molecules the emission was only calculated in LTE, using the parameters from CDMS \citep{Muller2001,Muller2005}.

For CH$_3$OH, H$^{13}$CO$^+$, CN and the other molecules we ran models to constrain the
upper limits. For CH$_3$OH, the same abundance profiles as for H$_2$CO
were taken because CH$_3$OH is expected to be cospatial with H$_2$CO when
they are both formed through solid-state chemistry. For the
H$^{13}$CO$^+$ abundance the initial CO abundance was taken and
multiplied with factors 10$^{-5}$--10$^{-11}$, as the HCO$^+$ is known
to form via CO through the H$_3^+$ + CO proton donation reaction and
is indeed observed to be strongly correlated with CO
 \citep{Jorgensen2004}. For the other molecules the same approach for the abundance as for H$^{13}$CO$^+$ was used.

\subsection{Results}
\label{sect:modelresults}

\begin{figure}[!ht]
\begin{center}
\includegraphics[scale=0.6]{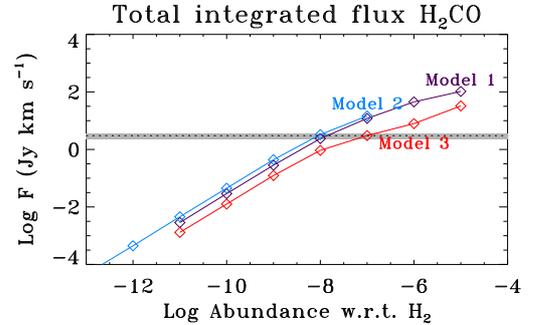}
\caption{Results of H$_2$CO abundance models: total flux integrated over the emission rectangle of the observations for different abundances for model 1 (purple), model 2 (blue), and model 3 (red). The fractional abundances with respect to CO for model 2 have been multiplied with 10$^{-4}$ for easier comparison as the CO/H$_2$ abundance is typically 10$^{-4}$. The dotted line indicates the measured observed flux, and the gray bar indicates the error on this value based on the flux calibration uncertainty.}
\label{fig:trends}
\end{center}
\end{figure}

\begin{figure}[!ht]
\begin{center}
\subfigure{\includegraphics[trim=0 30 0 0,scale=0.4]{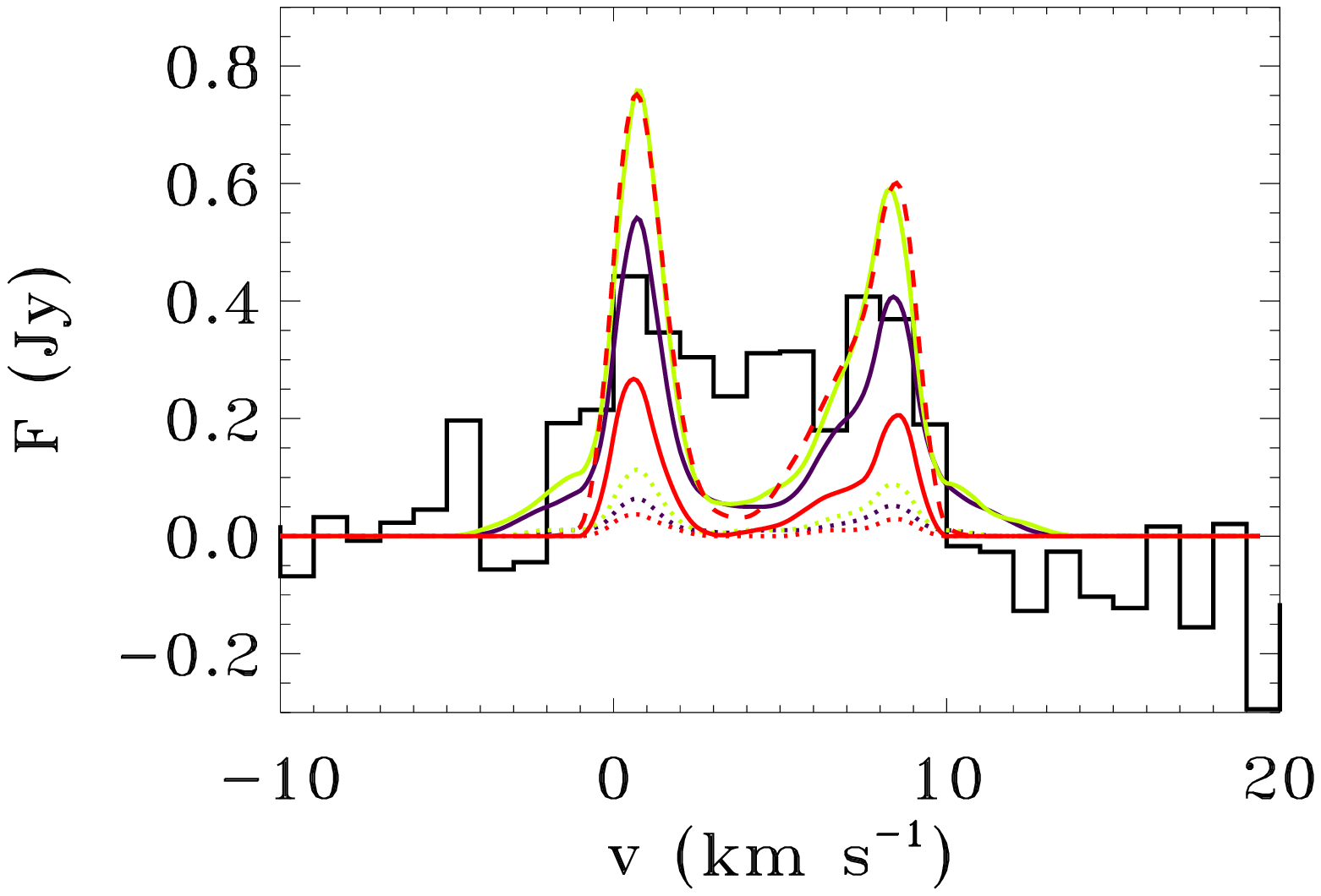}}
\subfigure{\includegraphics[trim=0 0 0 30,scale=0.4]{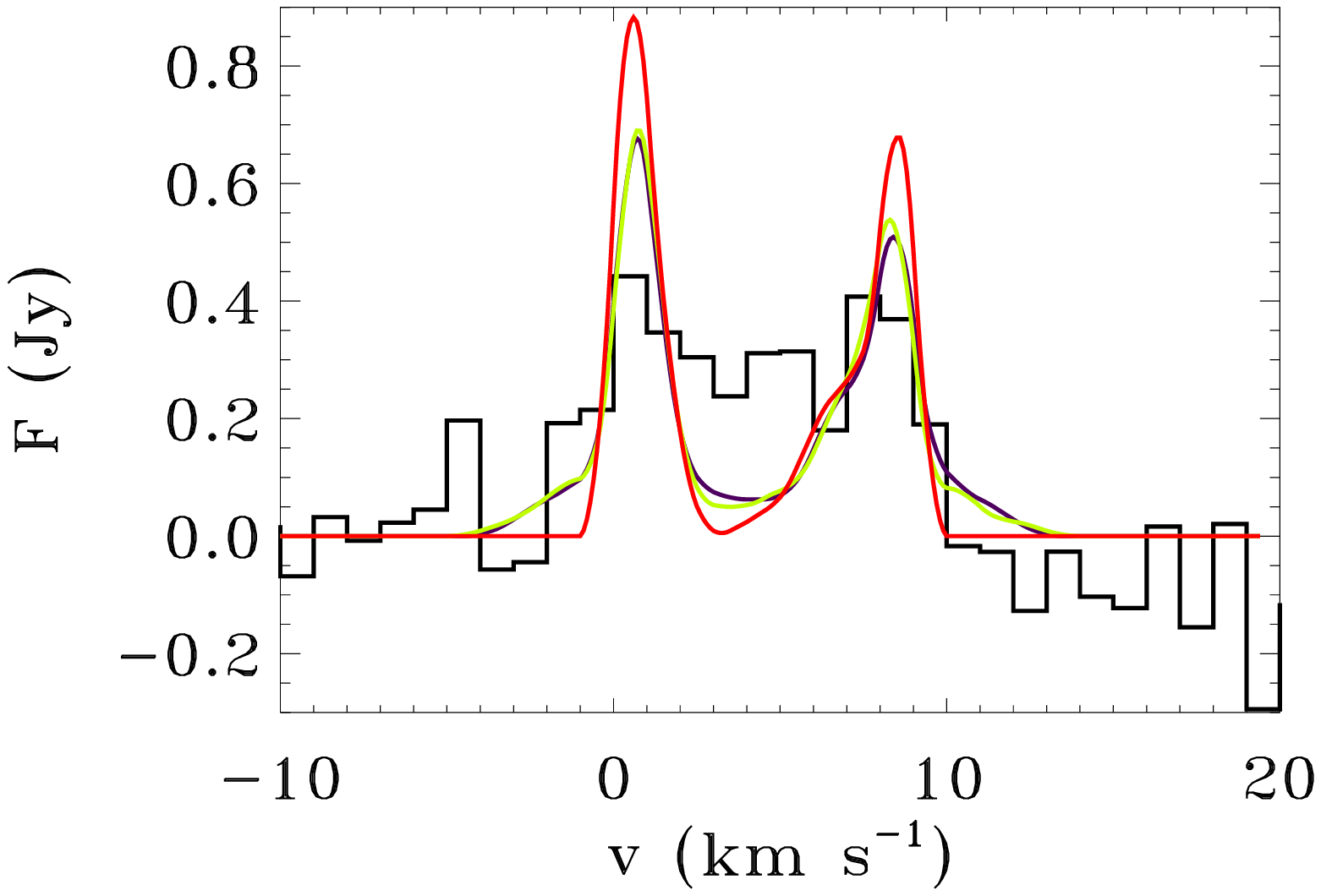}}
\caption{Results of H$_2$CO abundance models: spectra integrated over the emission rectangle of the observations for different abundances for model 1 (purple), model 2 (green), and model 3 (red) for abundances 10$^{-7}$ (dashed), 10$^{-8}$ (solid), and 10$^{-9}$ (dotted). The black spectrum represents the observational data. Abundances for model 2 are multiplied with 10$^{-4}$ to translate the abundance w.r.t. CO to H$_2$. The top figure shows the original models. The bottom figure shows the model spectra scaled to match the total flux of the observations.}
\label{fig:modelspectra}
\end{center}
\end{figure}

\begin{figure}[!ht]
\begin{center}
\includegraphics[scale=0.45,trim=0 0 0 0]{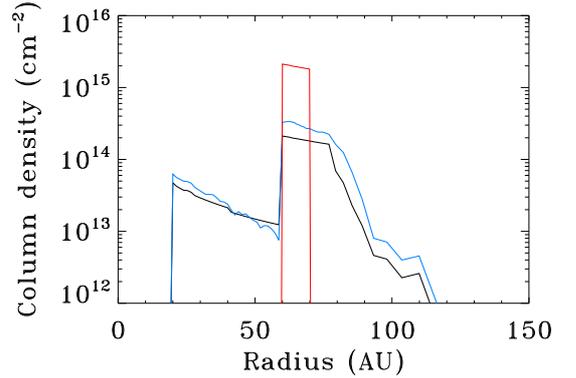}
\caption{Column density profiles of H$_2$CO for the best abundance fits for models 1 (black), 2 (blue), and 3 (red). Abundances are 10$^{-8}$ for model 1, 10$^{-4}$ w.r.t. CO for model 2 and 10$^{-7}$ for model 3.}
\label{fig:coldens}
\end{center}
\end{figure}

The model results for H$_2$CO are presented in Figure \ref{fig:trends}.
Integrated fluxes were computed by summing the model
fluxes over the same rectangular region as the H$_2$CO observations, after subtracting the continuum.  Figure \ref{fig:trends} presents the model
fluxes as a function of H$_2$CO abundance for the three trial abundance
structures.  Model 1 with constant H$_2$CO abundance $\sim$10$^{-8 \pm 0.15}$
with respect to H$_2$ or model 2 with abundance $\sim$10$^{-4}$ with
respect to CO reproduce the total observed flux well within the error bar. There is little
difference between model 1 and model 2 except for a factor of 10$^4$ that has
been taken into account in Figure \ref{fig:trends}. This is expected
because most of the CO abundance in the defined region is
10$^{-4}$ with respect to H$_2$. Model 3 requires an abundance
of 10$^{-8}$ higher by a factor 3 in the 60--70 AU ring to give the same integrated
flux. Note that the LIME model fluxes obtained with non-LTE
calculations are only about 25\% lower than the LIME models in LTE due
to the high densities in the disk.

The spectra (Figure \ref{fig:modelspectra}) confirm that the best match for the flux for
model 1 is for an abundance of $\sim$10$^{-8}$, although the peaks at the
highest velocities are up to twice as high as the data. The slight asymmetry in these model spectra
is caused by the spatial integration over a rectangle, whereas the
disk has a position angle of 96$^{\circ}$. 
The line wings in models 1 and 2 originate from the emission from
radii $<$60 AU, which is missing in model 3, but the $S/N$ of the data is
too low to detect the difference. In the bottom panel of Figure \ref{fig:modelspectra} the spectra have been scaled so that the total flux exactly
matches that from the observations.The current models underproduce
the emission close to the central velocity ratios, suggesting that the abundance may be even more enhanced in the central southern part of the disk at these velocities and is not constant along the azimuthal direction of the semi-ring. It is also possible that there is enhanced H$_2$CO at larger radii than assumed here when the freeze-out zone is taken out, although this does not add a significant amount of emission at the central velocities. Calculation of an abundance model where the freeze-out zone is removed shows that this indeed increases the emission at central velocities, but the $S/N$ of the data is insufficient to confirm or exclude emission at larger radii. 
Overall, it is concluded that the abundance is $\sim10^{-8}$ compared
with H$_2$ within factors of a few.

\begin{figure*}[!ht]
\begin{center}
\includegraphics[scale=1]{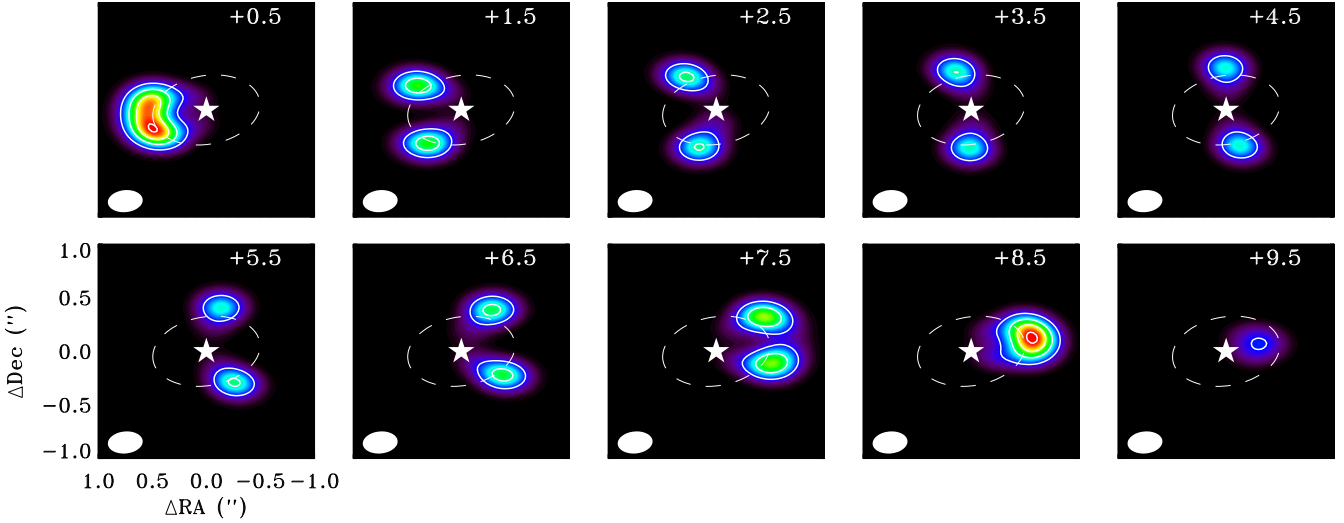}
\caption{Results of H$_2$CO abundance modeling: channelmap convolved with the ALMA beam of the observations for Model 1 at 10$^{-8}$ abundance. The color bar gives the flux scale in mJy beam$^{-1}$. The 60 AU radius is indicated with a dashed ellipse and the stellar position with a star. The white contours indicate the 20\%, 40\%, 60\%, 80\% and 100\% of the peak (peak is 212 mJy beam$^{-1}$). }
\label{fig:modelchannelmaps}
\end{center}
\end{figure*}

\begin{figure*}
\hspace{4cm}Model 1\hspace{3cm}Model 2\hspace{3cm}Model 3
\begin{center}
\subfigure{\includegraphics[scale=0.55, trim=40 0 40 0]{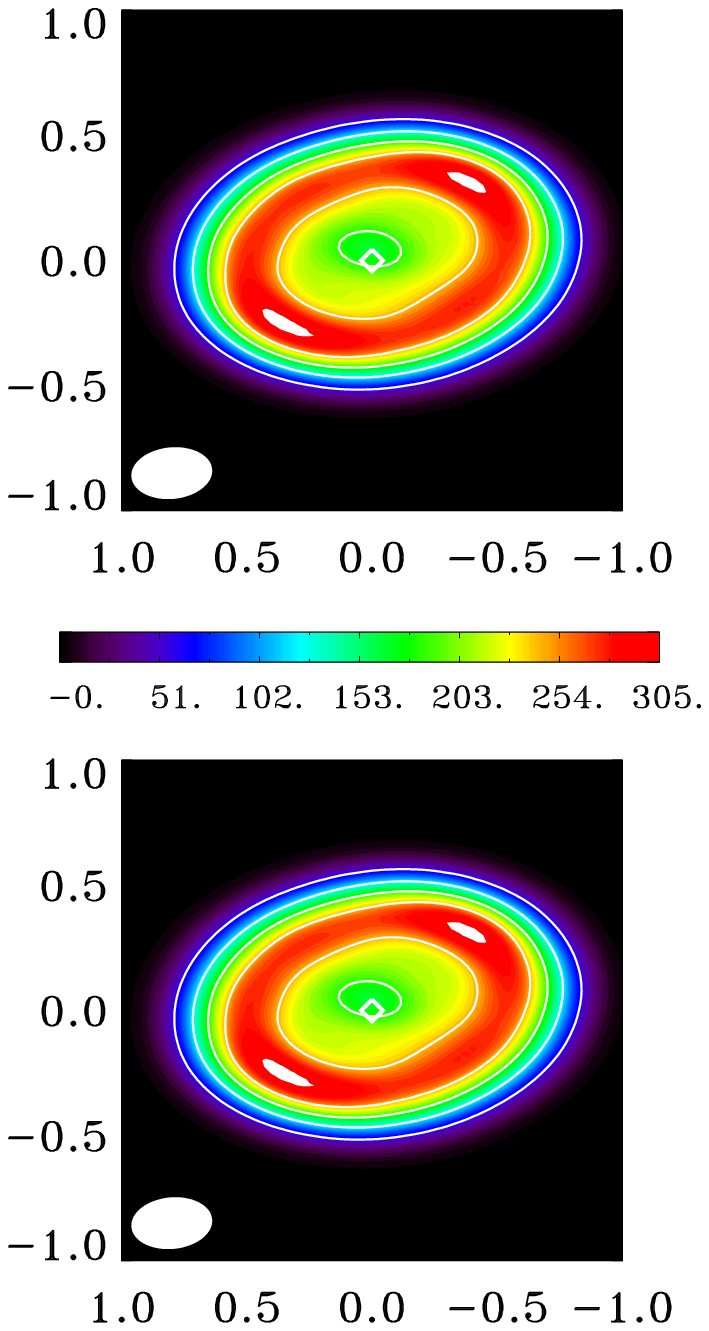}}
\subfigure{\includegraphics[scale=0.55, trim=40 0 40 0]{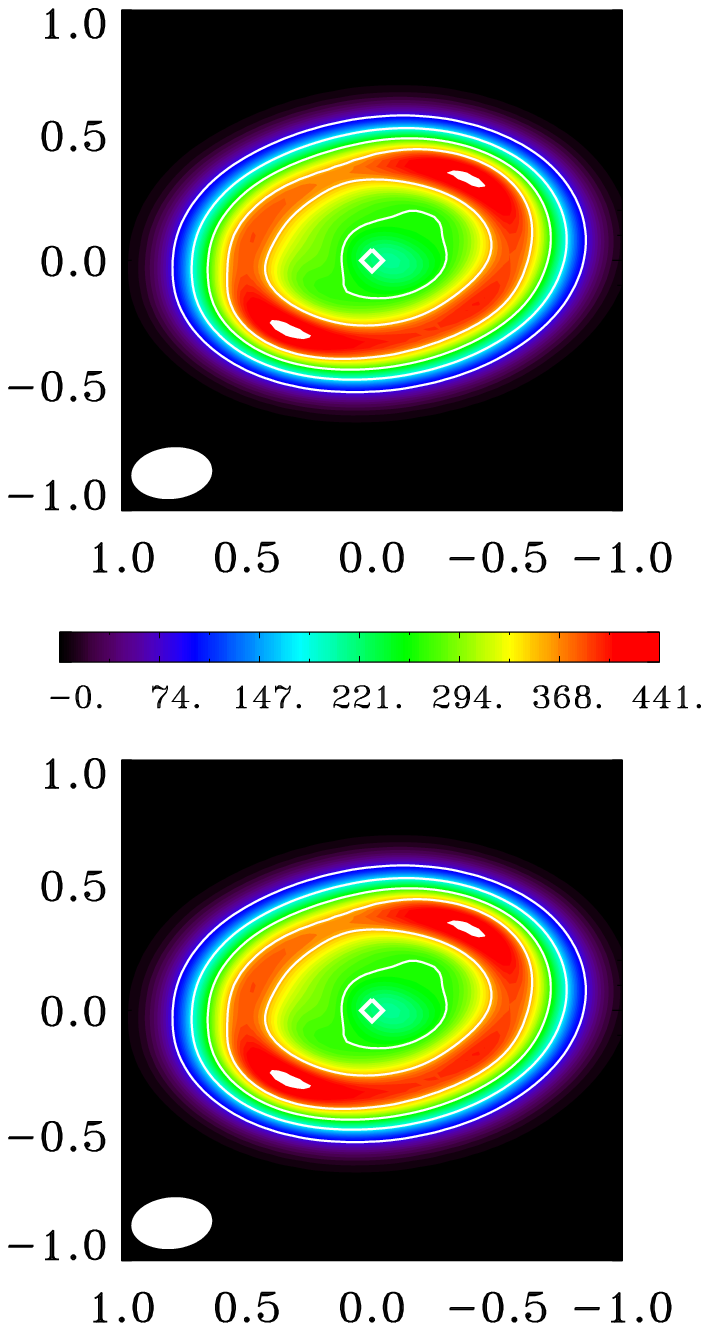}}
\subfigure{\includegraphics[scale=0.55, trim=40 0 40 0]{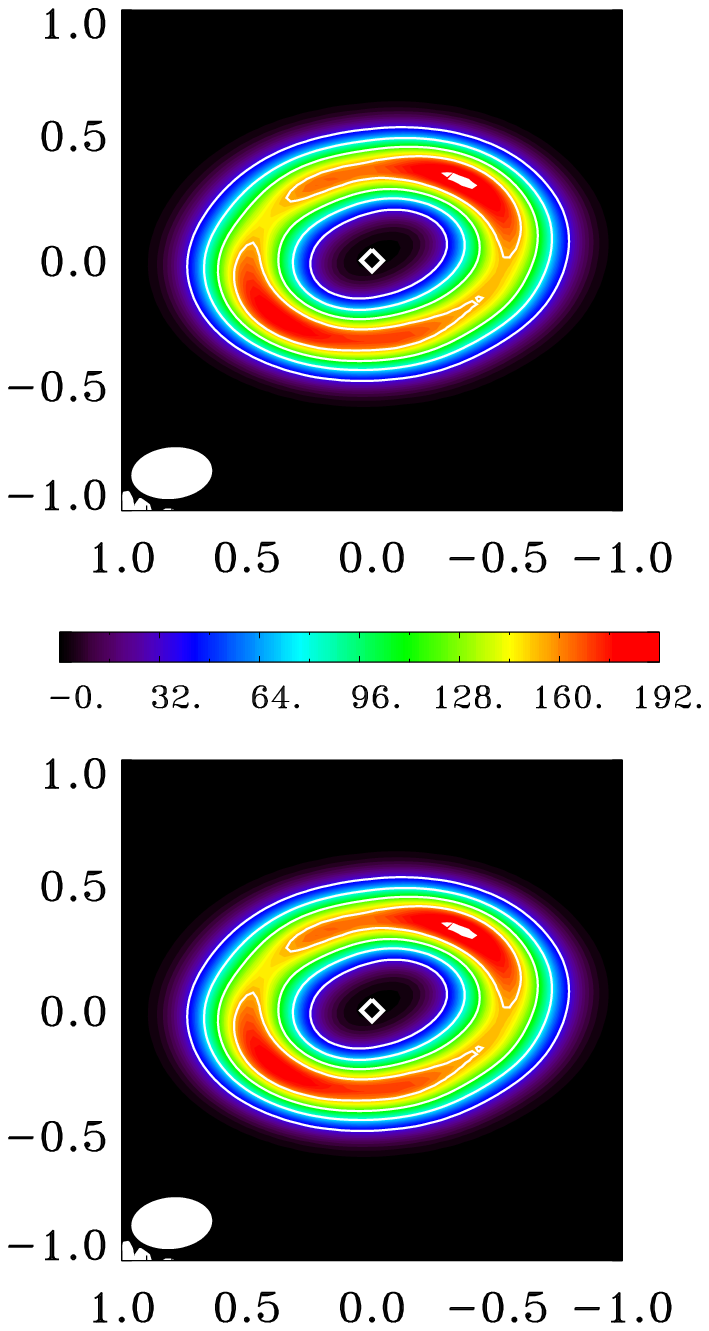}}\\
\caption{Results of H$_2$CO abundance models: Integrated intensity maps convolved with the ALMA beam for Model 1 at 10$^{-8}$ abundance (left), Model 2 at 10$^{-4}$ fractional abundance of CO (center) and Model 3 at 10$^{-8}$ abundance (right). The color bar gives the flux scale in mJy beam$^{-1}$. The white contours indicate the 20\%, 40\%, 60\%, 80\% and 100\% of the peak. The top panel gives the result for the non-altered model image, the bottom panel gives the result with a correction for the continuum extinction. }
\label{fig:modelmomentmaps}
\end{center}
\end{figure*}

The final comparison between models and data is made by comparing images. 
To produce images from the model output cubes the images were convolved
with the ALMA beam of the observations (0.31''$\times$0.21'', PA
96$^{\circ}$). Similar as in \citet{BruderervdM}, the result of the model images convolved to the ALMA beam was compared with the result of
simulated ALMA observations. An alternative method is to convert the model images to (u, v)-data according to the observed (u, v)-spacing using the CASA software and reduce them in the same way as the observations. Because of the good (u, v)-coverage of our observations, the two approaches do not differ measurably within the uncertainty errors.
Figures \ref{fig:modelchannelmaps} and \ref{fig:modelmomentmaps} indicate that the three models have a similar ring-like structure as the observations, apart from the
emission in the north that is lacking in the data. The differences
between the models are best seen in the velocity channel maps: models 1
and 2 still show some emission within the ring at 30--60 AU, which is
higher for model 1 than for model 2 because the CO abundance is
somewhat lower between 40 and 60 AU. Model 3 does not show any
emission within the ring by design.  

A possible explanation for the missing H$_2$CO emission at the peak of the dust continuum is that the dust is not entirely optically thin. The optical depth $\tau_{\rm d}$ was calculated as $\tau_{\rm d}\sim 0.43$ \citep{vanderMarel2013} averaged over the continuum region. If we assume all the H$_2$CO emission $I_{\rm line}$ to originate from behind the dust, the resulting intensity is
\begin{equation}
I_{\nu} = B_{\nu}(T_{\rm dust})\cdot(1-e^{-\tau_{d}}) + e^{-\tau_{\rm d}}I_{\rm line}
.\end{equation}
The first term is the measured dust intensity, which is subtracted from the H$_2$CO data. The second term indicates a reduction of the line intensity by continuum extinction. This extinction was calculated by multiplying the model output with the exponent of a 2D $\tau_{\rm d}$ profile, where $\tau_{\rm d}$ is following the continuum emission profile of the observed dust trap area. The maximum $\tau_{\rm d}$ was taken as 0.86, recovering the averaged opacity over this area. This correction represents an upper limit of the continuum extinction effect. The result is shown in the bottom panels of Figure \ref{fig:modelmomentmaps}. The continuum extinction decreases the H$_2$CO emission in the south by more than a factor 2, while some emission between the star and the dust continuum remains (for models 1 and 2). Although the strength of this emission compared with the peaks at the edges is still lower than in the observations, the model image is now more consistent with the observations. The total integrated flux is only 10\% lower because the rectangle used for the spatial integration did not cover most of the dust continuum. Both the $S/N$ of the line data and the unknown mixing of the gas and dust prevent a more detailed analysis of this problem, but it is clear that the continuum extinction cannot be neglected. The observations show a decrease of a factor 4 between the east and west limbs compared with the emission at the location of the dust peak (see Figure 1), which requires a $\tau_{\rm d}$ of at least 1.4. Submillimeter imaging at longer wavelengths is required to measure the dust optical depth more accurately.

Overall, the conclusion is that
the current data can constrain the H$_2$CO abundance in the warm gas
where H$_2$CO is thought to reside to $\sim 10^{-8}$, but the current
$S/N$ is too low to distinguish the different assumptions on the
radial distribution of H$_2$CO. However, the three models make
different predictions for the distribution, which we expect to
see more clearly in future higher $S/N$ ALMA data.

\begin{figure}[!ht]
\begin{center}
\subfigure{\includegraphics[scale=0.6]{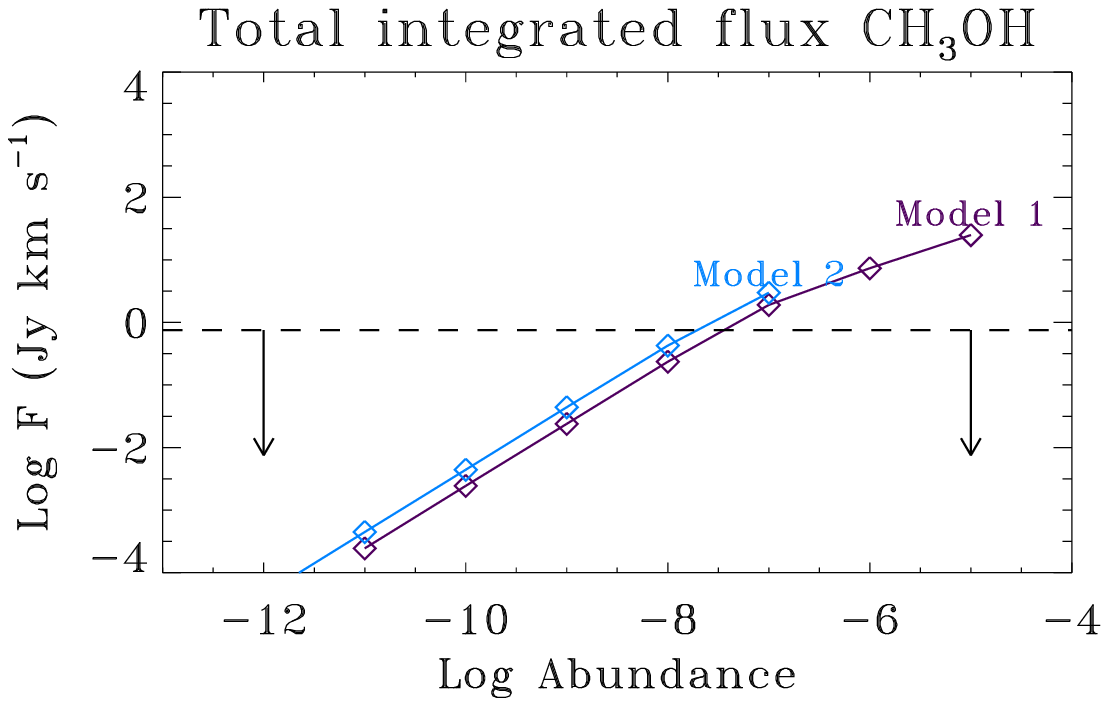}}
\subfigure{\includegraphics[scale=0.6]{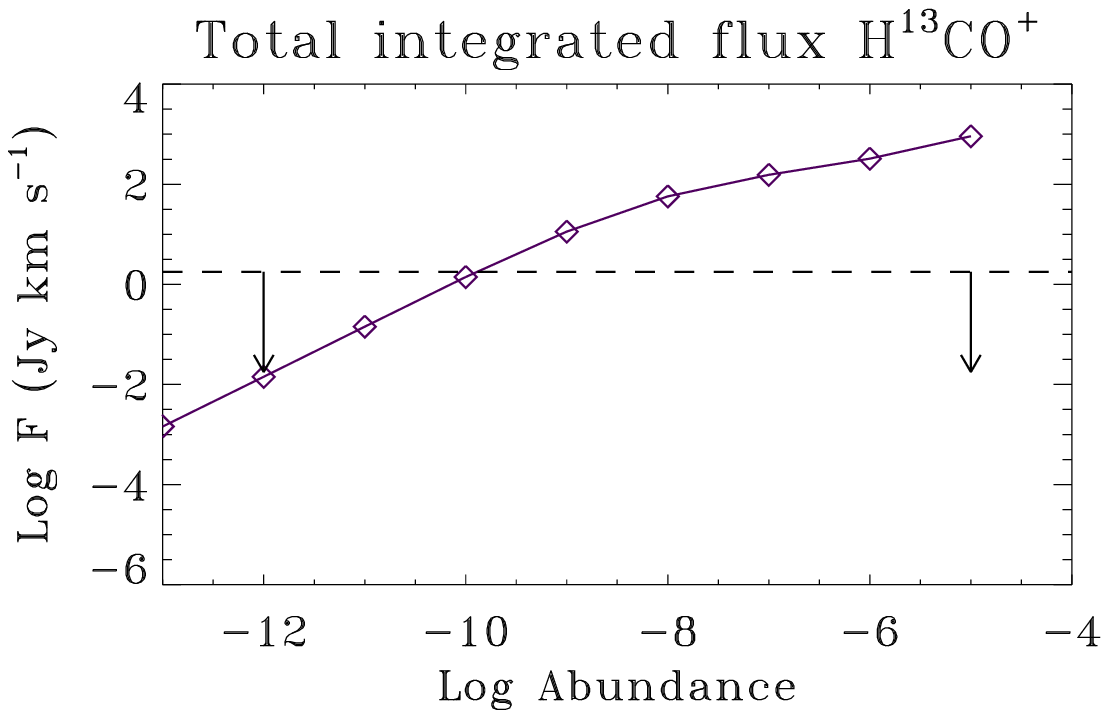}}
\subfigure{\includegraphics[scale=0.6]{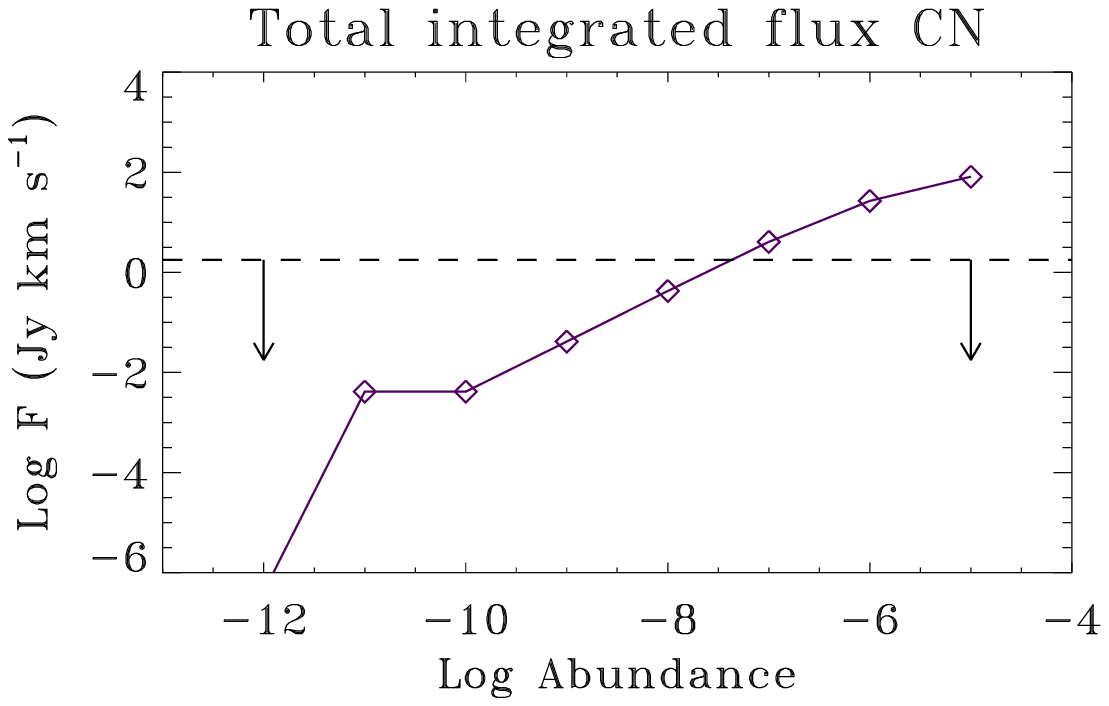}}
\caption{Model results for the integrated fluxes found for CH$_3$OH, H$^{13}$CO$^+$ and CN in the empirical models. The dashed line indicates the upper limit of the integrated intensity, assuming the molecule is cospatial with the H$_2$CO (CH$_3$OH) or CO (H$^{13}$CO$^+$ and CN). Abundances were calculated with respect to the CO abundance, but in these plots are multiplied with 10$^{-4}$ to translate the abundance w.r.t. CO to H$_2$.}
\label{fig:othermols}
\end{center}
\end{figure}

The total model fluxes for the CH$_3$OH, H$^{13}$CO$^+$ and CN
are compared with the derived upper limits in Figure
\ref{fig:othermols}, where all abundances have been multiplied with 10$^{-4}$ to translate the abundance w.r.t. CO to H$_2$. The other targeted molecules with upper limits were compared in the same way (plots not displayed here). For the CH$_3$OH lines, the upper limit on the
integrated flux is consistent with an abundance limit in model 1 of
$<3\cdot10^{-8}$, thus H$_2$CO/CH$_3$OH $>$ 0.3. For H$^{13}$CO$^+$, the
upper limit sets the abundance at $<10^{-6}$ with respect to CO. This
indicates an HCO$^+$/CO abundance ratio of $<10^{-4}$, or an absolute
abundance HCO$^+$/H$_2$ of $<10^{-8}$. The CN emission is poorly
constrained: the upper limit sets the CN/CO abundance at $<5\cdot10^{-4}$
or CN/H$_2 < 5\cdot10^{-8}$. The reason is the low Einstein A coefficient of
this particular transition, which is almost three orders of magnitude
lower than that of the other transitions in this study. 

\begin{table}[h]
\small
  \caption{Derived abundance limits w.r.t. H$_2$.}
  \label{tbl:abundances}
  \begin{tabular}{ll}
    \hline
    Molecule & Abundance\\
    \hline
        H$_2$CO&$1\cdot10^{-8 }$\\
        CH$_3$OH&$<3\cdot10^{-8 }$\\
        H$^{13}$CO$^+$&$<1\cdot10^{-10}$\\
        CN&$<5\cdot10^{-8 }$\\
        $^{34}$SO$_2$&$<1\cdot10^{-8 }$\\
        C$^{34}$S&$<1\cdot10^{-9 }$\\
        HNCO&$<3\cdot10^{-9 }$\\
        c-C$_3$H$_2$&$<3\cdot10^{-9 }$\\
        N$_2$D$^+$&$<1\cdot10^{-10 }$\\
        D$_2$O&$<1\cdot10^{-9 }$\\      
    \hline
  \end{tabular}
\end{table}

All derived absolute abundance limits are given in Table \ref{tbl:abundances}. For $^{34}$SO$_2$ and C$^{34}$S we found absolute abundances of $<10^{-8}$ and $<10^{-9}$ respectively, corresponding to abundances of $<2\cdot10^{-7}$ and $<2\cdot10^{-8}$ for the main isotopes, assuming an elemental abundance ratio of sulfur of $^{32}$S/$^{34}$S of 24 \citep{WilsonRood1994}. For the deuterated molecules N$_2$D$^+$ and D$_2$O the isotope ratio D/H in these molecules is not known, therefore no upper limits on N$_2$H$^+$ or H$_2$O can be obtained.

\section{Discussion}
\subsection{Origin of the H$_2$CO emission}
H$_2$CO can be formed efficiently in the ice phase by hydrogenation of solid CO, as shown in laboratory experiments, which can be followed by the formation of CH$_3$OH \citep{Hiraoka2002,Watanabe2002,Watanabe2004,Hidaka2004,Fuchs2009}:
\begin{equation}
{\rm CO} \rightarrow {\rm HCO} \rightarrow {\rm H}_2{\rm CO}  \rightarrow {\rm H}_2{\rm COH}  \rightarrow {\rm CH}_3{\rm OH}\\
.\end{equation}
Because CO ice is highly abundant in the cold dusty regions in clouds
and disks, this formation route is commonly assumed as the origin of
gas-phase H$_2$CO, following thermal or nonthermal desorption. In
that case, CH$_3$OH is expected to have a similar or higher abundance
than H$_2$CO because they are formed along the same sequence
 \citep{Tielens1982,vanderTak2000,Cuppen2009}. If all ices are thermally
desorbed, gas-phase abundances can be as high as $\sim 10^{-6}-10^{-5}$.

There is no known efficient gas-phase chemistry for the formation of
CH$_3$OH \citep{Geppert2006,Garrod2006FD}, whereas H$_2$CO can also be formed
rather efficiently in the gas phase. At low temperatures, the reaction
\begin{equation}
{\rm CH}_3 + {\rm O} \rightarrow {\rm H}_2{\rm CO} + {\rm H}
\end{equation}
dominates its formation, whereas it is mainly destroyed through
reactions with ions such as HCO$^+$and H$_3$O$^+$. These reactions lead to
H$_3$CO$^+$, which cycles back to H$_2$CO through dissociative recombination
\begin{equation}
{\rm H}_3{\rm CO}^++ e^- \rightarrow {\rm H}_2{\rm CO} + {\rm H}
.\end{equation}
However, since the branching ratio to H$_2$CO is only 0.3
 \citep{Hamberg2007}, the ion reactions lead to a net destruction of
H$_2$CO. Typical dark-cloud gas-phase model abundances are a
few$\times 10^{-8}$ relative to H$_2$ \citep{McElroy2013}. At low dust
temperatures ($<$ 60 K), H$_2$CO can freeze out after formation.

At higher temperatures in the presence of abundant H$_2$O ($T>100$~K),
formaldehyde can also form through gas-phase reactions such as
CH$_2^+$+H$_2$O $\rightarrow$ H$_3$CO$^+$ + H followed by dissociative recombination.
Above a few 100 K, formaldehyde is destroyed efficiently through H + H$_2$CO + 1380 K $\rightarrow$ HCO + H$_2$.

H$_2$CO has been detected in warm protostellar cores together with
CH$_3$OH. Typical H$_2$CO abundances in the warm gas are
3$\sim10^{-7}$ with respect to hydrogen, whereas CH$_3$OH has an abundance higher by a
factor of 5 (H$_2$CO/CH$_3$OH$\approx$0.2)
 \citep{Bisschop2007}, suggesting a main formation route through solid-state chemistry followed by sublimation. Recent observations of these molecules in the Horsehead PDR and core show a higher abundance ratio H$_2$CO/CH$_3$OH of 1--2, which cannot be produced by pure gas-phase chemistry \citep{Guzman2013}.  H$_2$CO has been detected in
several protoplanetary disks \citep{Dutrey1997,Aikawa2003,Thi2004,Oberg2010,Oberg2011,Qi2013}, which
has usually been interpreted through the solid state chemical path,
even though gas-phase CH$_3$OH has not been detected in any of these
disks due to sensitivity limitations. For high enough sensitivity the
combination of H$_2$CO and CH$_3$OH observations would clearly
distinguish between the ice-phase and gas-phase chemistry. For the disk around HD163296, the rotational temperature of H$_2$CO
was measured to be $<$30 K by fitting the emission of several
transitions, and its abundance was found to be enhanced outside the CO
`snowline' at 20 K, suggesting a solid-state route \citep{Qi2013}.

The measured H$_2$CO/CH$_3$OH ratio limit for Oph IRS 48 of $>$0.3 is
similar to the limit found for LkCa 15 \citep{Thi2004} and only
slightly higher than that found in warm protostars. This low limit
therefore does not exclude the solid-state chemistry formation route
but is also consistent with a gas-phase chemistry contribution to
H$_2$CO. However, the H$_2$CO abundance of 10$^{-8}$ is much higher
than the detections and upper limits of the disk-averaged abundances
in \citet{Thi2004}, which are typically $<10^{-11}$.  Most of this
difference stems from the fact that the disks in \citet{Thi2004} are
two orders of magnitude more massive and also colder, with a large
portion of the CO and related molecules frozen out. 
This cold zone is lacking in the IRS 48 disk, and our analysis has
already removed the zones in which H$_2$CO is photodissociated or
frozen out. Nevertheless, as shown below, the H$_2$CO emission is
unusually strong in IRS 48 for its disk mass, hinting at an increased
richness of chemistry in transitional disks due to the directly irradiated inner rim at the edge of the gas gap, as predicted \citep{Cleeves2011}.

Another interesting aspect of the H$_2$CO emission is the 
azimuthal asymmetry in IRS 48, which happens to be similar although
not exactly cospatial to the dust continuum asymmetry. The dust
asymmetry was modeled \citep{vanderMarel2013} as a dust trap caused by a
vortex in the gas distribution with an overdensity of a factor 3, and
it is possible that the increased emission of H$_2$CO at this
location is in fact tracing this overdensity in the gas. The image
quality and $S/N$ of the observations is insufficient to confirm
this claim, however, and no overdensity was observed in the $^{12}$CO 6--5 emission \citep{BruderervdM}. Moreover, the H$_2$CO emission does not peak exactly at the
peak of the dust continuum emission: in fact, the H$_2$CO emission is
decreased at the continuum peak, as discussed in Sect. \ref{sect:modelresults} (see Figure \ref{fig:momentmap}). It
is possible that this decrease is caused by the high optical depth of the
millimeter dust, which absorbs the line emission of the optically thin gas
all the way from the midplane (see Figure \ref{fig:modelmomentmaps} and Eq. 3). This cannot be confirmed because of the
unknown vertical mixing of the gas and dust. Finally, a nondetection of an overdensity in the gas is consistent with the dust-trapping scenario: a vortex in the gas may already have disappeared while the created dust asymmetry remains because the time scale of smoothing out a dust asymmetry is on the order of several Myr \citep{Birnstiel2013}. 

Another possibility for the increased H$_2$CO emission in the south is
a local decrease in the temperature compared with the north. In that
case H$_2$CO is destroyed efficiently in the north through the H +
H$_2$CO + 1380 K $\rightarrow$ HCO + H$_2$ reaction, but not in the
south. This possibility is consistent with the reduced CO emission in
the south in both the rovibrational lines \citep{Brown2012a} and the
$^{12}$CO 6--5 lines \citep{BruderervdM}, where \citet{BruderervdM}
suggested the temperature decrease to be of a factor 3. This temperature decrease might be caused by UV shielding in the inner disk. Note that the millimeter dust that is
concentrated in the south does not provide sufficient cooling to
change the gas temperature significantly through gas-grain
collisions: the small dust grains have most the surface and gas-grain temperature exchange. If the temperature drops locally to between 20 and 30 K,
the H$_2$CO might even be an ice-phase product, but this is very
unlikely. A final possibility is that the increase of H$_2$CO is
caused by the increased dust density and dust collisions in this
region. If the grain-grain collisions are at high enough speed (which
depends on the turbulence in the disk), the ice molecules may become separated from the grains. However, the dust grains have reached temperatures
above the sublimation temperature of 60 K long before reaching the dust
trap, therefore this is not very likely within reasonable time scales.

\subsection{Comparison with chemical models}
The column density of H$_2$CO for the best-fitting model with abundance 10$^{-8}$ is $\sim10^{14}$ cm$^{-2}$ for radii$>$60 AU (Figure \ref{fig:coldens}), which is within an order of magnitude of the predictions of chemical models \citep{Semenov2011,Walsh2012,Walsh2013} for protoplanetary disks. For the 10$^{-7.5}$ abundance upper limit for CH$_3$OH we found a column density limit of $\sim2-4\cdot10^{14}$ cm$^{-2}$ for radii$>$60 AU, which is well above the numbers derived in chemical models \citep{Walsh2012,Vasyunin2011}, which are typically 10$^{10}$ cm$^{-2}$. 

The DALI-model \citep{BruderervdM} produces predictions
for HCO$^+$ and CN: the HCO$^+$ abundance peaks at 5$\cdot10^{-9}$, but only in a very thin upper layer of the disk where the CO + H$_2$ reaction operates. The disk-averaged abundance is much lower. The predicted integrated model intensity for H$^{13}$CO$^+$ is 0.02 Jy km s$^{-1}$, far below our detection limit of 1.8 Jy km s$^{-1}$. The peak abundance for CN is 3$\cdot10^{-7}$ with an integrated intensity of 1.1 Jy km s$^{-1}$, which is close to our upper limit. The other CN 6--5 lines at 680 GHz (outside the range
of the spectral window of our observations) have an Einstein
coefficient that is 2 orders of magnitude higher and should be readily
detectable with an integrated flux of 15 Jy km s$^{-1}$.

\subsection{Comparison of upper limits with other observations}

The HCO$^+$/CO ratio of $<10^{-4}$ (or HCO$^+$/H$_2<10^{-8}$) for
IRS~48 is consistent with values found in disks, protostellar regions
and dark clouds. The ratio is often found to be $\geq 10^{-5}$,
where the lower limit is due to the unknown optical depth of the
observed HCO$^+$ line \citep{Thi2004}. Our H$^{13}$CO$^+$ line does not
suffer from this problem, but our inferred HCO$^+$
abundance is not more stringent because of the very low gas mass of IRS
48.  For similar abundances, all measured fluxes would be a factor of
10--100 lower than for other full disks.

HCO$^+$ is produced by the gas-phase reaction H$_3^+$ + CO
$\rightarrow$ HCO$^+$ + H$_2$, which relates the abundance directly to
ionization, because the parent molecule H$_3^+$ is formed efficiently
through cosmic-ray ionization. Our abundance cannot set a strong
limit on the ionization fraction in the disk. The ionization fraction
in disks is important because the magneto-rotational instability
(MRI), believed to drive viscous accretion, requires ionization to
couple the magnetic field to the gas \citep{Gammie1996}. Insufficient
ionization may suppress the MRI and create a so-called dead zone
that can create dust traps at its edge where the dust grains will
gather \citep{Regaly2012}. However, the ionization needs to be lower
than about 10$^{-12}$ to induce a dead zone \citep{Ilgner2006}. On the
other hand, recent models suggest that cosmic rays may be excluded
altogether from disks around slightly lower-mass stars
 \citep{Cleeves2013}. A detection of HCO$^+$ at the level suggested by
our models would provide direct proof of the presence of cosmic rays
that ionize H$_2$ at a rate of $\sim 5\times 10^{-17}$ s$^{-1}$.

The CN upper limit is difficult to compare with literature values because
our upper limit is very high due to the low Einstein A
coefficient. Literature values give derived abundances \citep{Thi2004}
of $\sim10^{-10}$, three orders of magnitude lower than our upper
limit, which is again caused by the low gas mass of our disk. As noted
above, the full chemical model by \citet{Bruderer2013} suggests
abundances very close to our inferred upper limits. The CN/HCN ratio
is a related tracer for photodissociation in the upper layers and at
the rim of the outer disk: a high ratio indicates a strong UV field, since CN is produced by radical
reactions with atomic C and N (in the upper layers) and by
photodissociation of HCN, whereas CN cannot easily be
photodissociated itself \citep{Bergin2003,vanDishoeck2006}. The CN/HCN
ratio is generally found to be higher in disks around the hotter
Herbig stars. HCN observations are required to measure this ratio for
IRS 48.

Several of the other targeted molecules have been detected towards cores and protostars, such as $^{34}$SO$_2$ \citep{Persson2012}, N$_2$D$^+$ \citep{Emprechtinger2009} and HNCO \citep{Bisschop2007}. N$_2$D$^+$ can be used as a deuteration tracer in combination with N$_2$H$^+$, and therefore a tracer of temperature evolution, but this line was not within our spectral setup. The H$_2$CO/HNCO abundance limit of $>$0.3 as derived for IRS 48 is rather conservative compared with values in cores of $\sim$10 \citep{Bisschop2007}, but the N-bearing molecules are weak in IRS 48. The c-C$_3$H$_2$ molecule was recently detected for the first time in the HD163296 disk \citep{Qi2013_c3h2}, and their derived column density of $\sim10^{12}$ cm$^{-2}$ is well below our observed limit for IRS 48 of 10$^{13}$ cm$^{-2}$.

\subsection{Predictions of line strengths of other transitions}
\label{otherlines}
\begin{figure*}[!ht]
\begin{center}
\subfigure{\includegraphics[trim=0 20 0 0,scale=0.4]{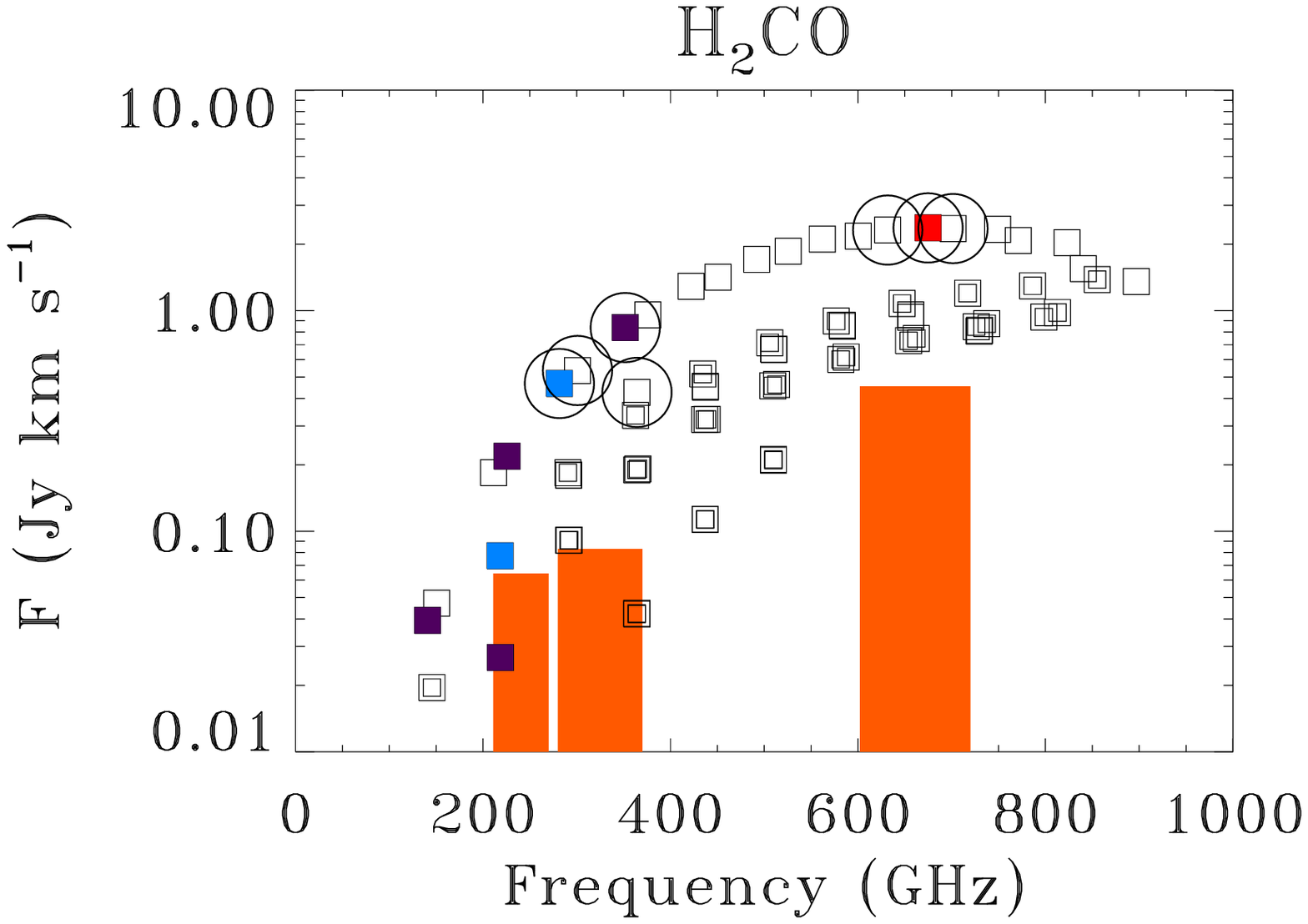}}
\subfigure{\includegraphics[trim=0 20 0 0,scale=0.4]{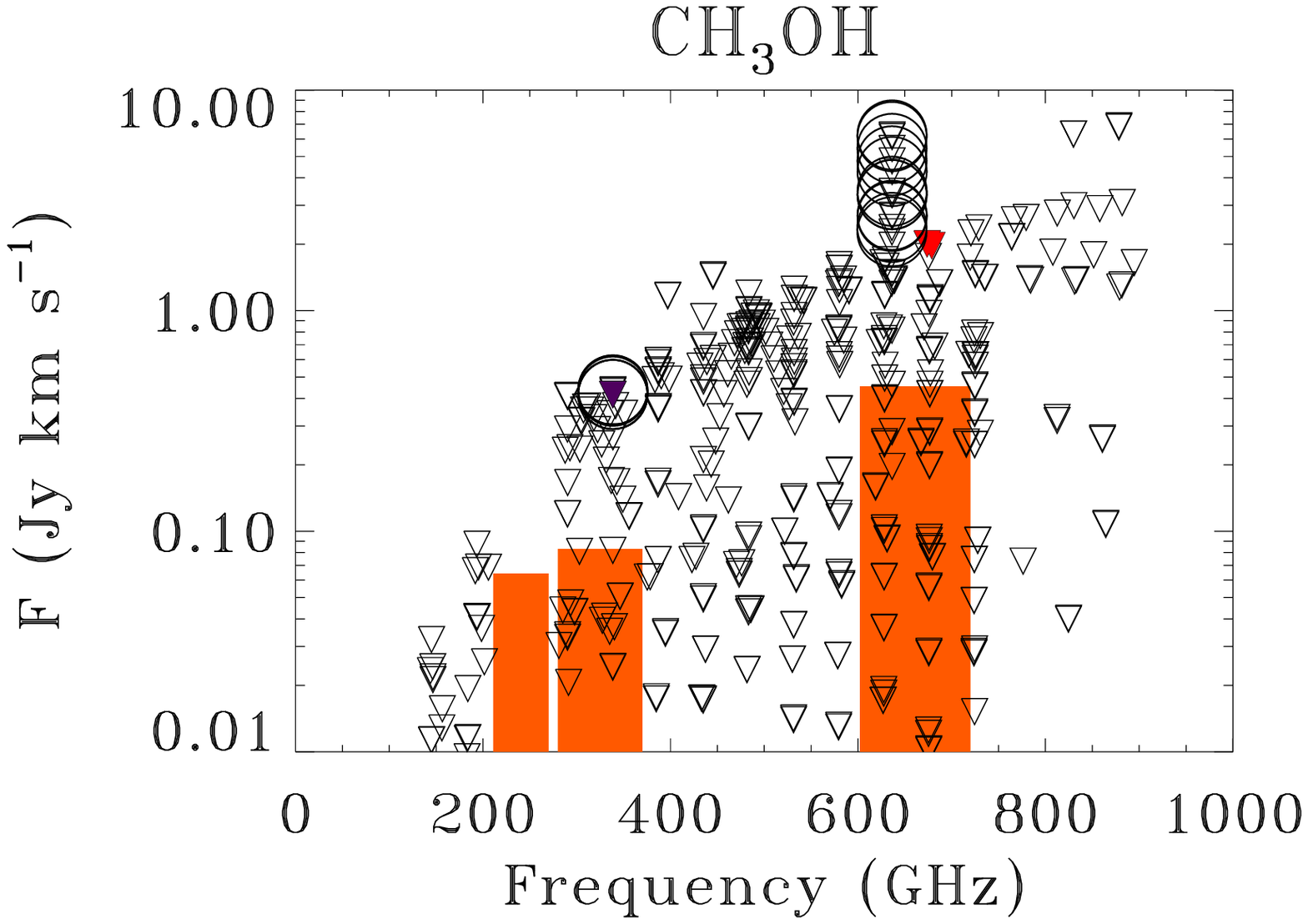}}\\
\subfigure{\includegraphics[trim=0 0 0 20,scale=0.4]{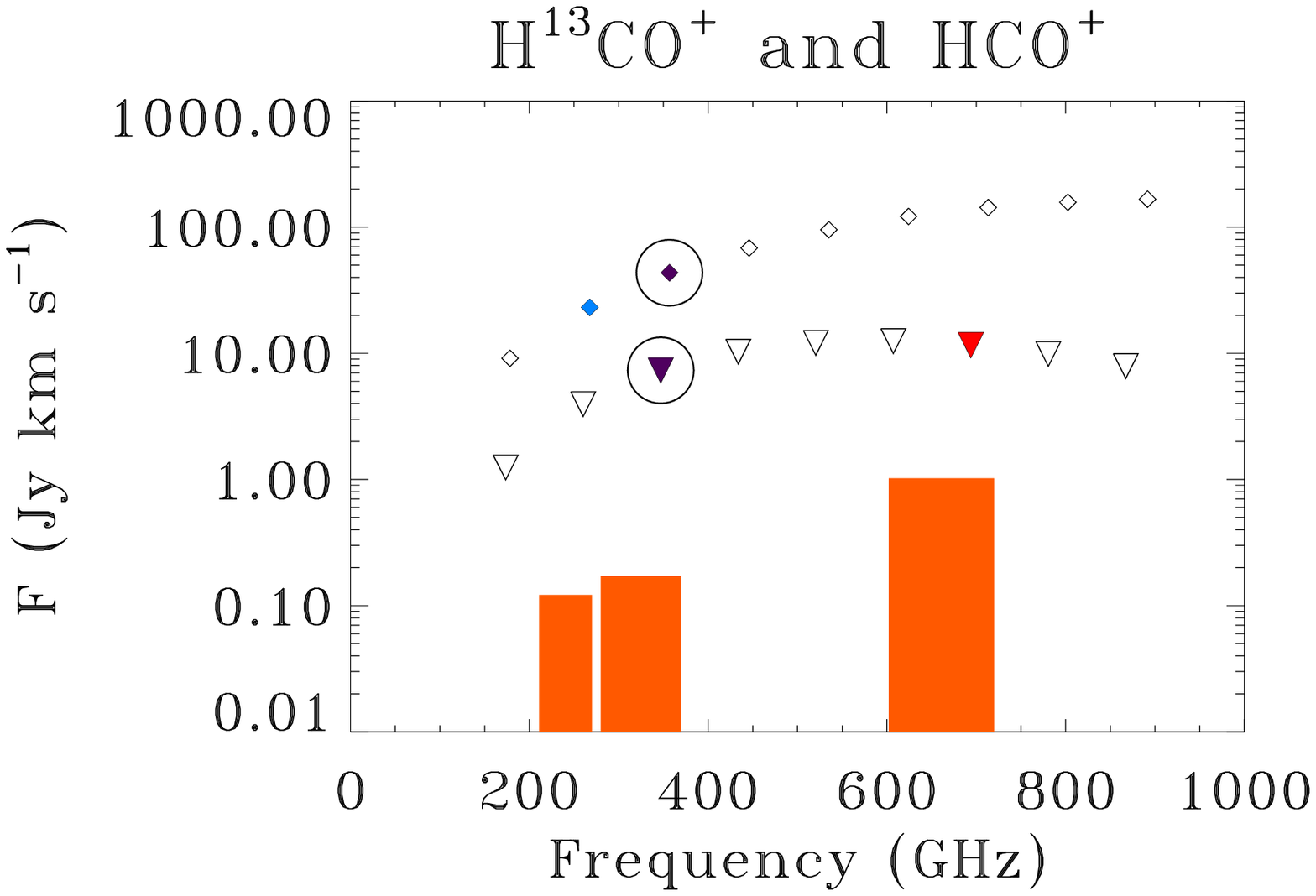}}
\subfigure{\includegraphics[trim=0 0 0 20,scale=0.4]{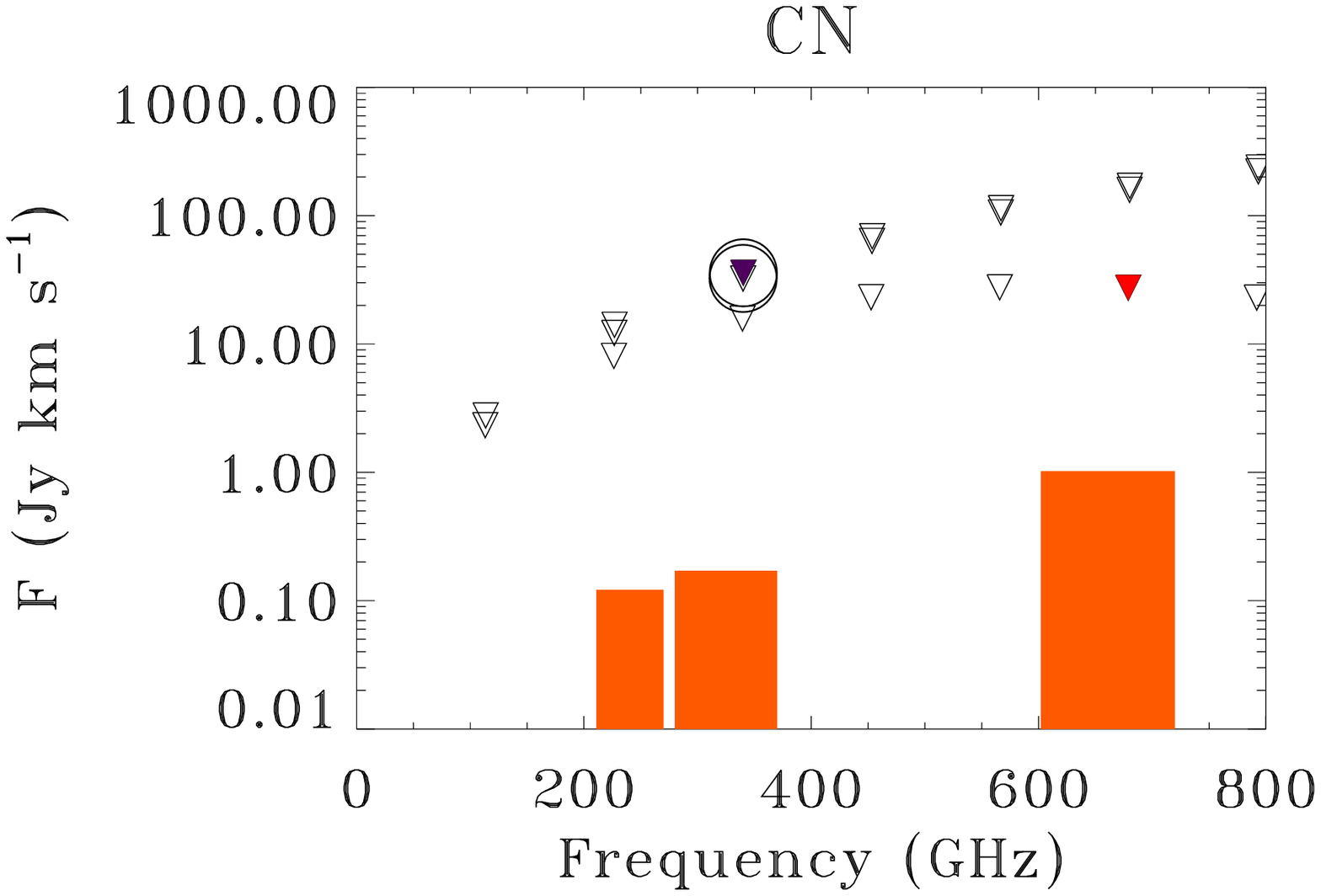}}
\caption{Model predictions for the integrated fluxes for H$_2$CO (both ortho and para lines, assuming an ortho/para ratio of 3), and upper limits for 
  A-CH$_3$OH, H$^{13}$CO$^+$, HCO$^+$ (triangles and diamonds,
  respectively), and CN based on our best-fit models for detections and
  upper limits. The boxes show the 3$\sigma$ upper limit of the ALMA
  sensitivity for Band 6 (230 GHz), Band 7 (345 GHz), and Band 9 (690
  GHz), integrated over the line profile for one hour integration in
  the full array. The targeted lines in observations are indicated in red (this study), blue ({\"O}berg et al. 2010,2011), and purple (Thi et al. 2004). The lines that have the best potential for observation compared with the ALMA sensitivity are encircled.}
\label{fig:othertrans}
\end{center}
\end{figure*}

Most observations of molecules in disks present only intensities, but
do not derive abundances, since a physical disk model is generally not
available. To compare our observations with these data, we
calculated the expected disk-integrated fluxes for other transitions,
using our physical model and derived abundances or upper limits. These
results are presented in Figure \ref{fig:othertrans}. The ALMA
sensitivity limits for Band 6, Band 7, and Band 9 (230, 345 and 690
GHz) for the ALMA full array of 54 antennas for 1 hour integration are
included to investigate which lines provide the best constraints for
future observations.

Figure \ref{fig:othertrans} shows the potential for future observations
of IRS 48, indicating that with full ALMA much better upper limits or
detections can be reached for all of these species in just one hour of integration by choosing the
correct lines. The H$_2$CO flux can be measured with much better $S/N$ and
there is a wide range of abundances that can be tested. Lines in Band 7 (345 GHz) generally have the highest potential. 

For H$_2$CO, lines that have been targeted most often are the para
H$_2$CO 3(0,3)-2(0,2) and 3(2,2)-2(2,1) transitions at 218.22 and 218.47
GHz, as well as the ortho-H$_2$CO 4(1,4)-3(1,3) line at 281.53 GHz
 \citep{Oberg2010,Oberg2011}.  Comparing the
observed fluxes with our predictions for these lines ($\sim$ 0.3 Jy km
s$^{-1}$) shows that the H$_2$CO emission of IRS 48 is quite similar
to that found in other disks. This is remarkable because of the low disk
mass of IRS 48. 

CH$_3$OH has not been detected in other disks to date. Upper limits
derived for a few disks \citep{Thi2004} give abundances $<10^{-10}$,
below our limits for IRS 48. The fact that our observed H$_2$CO fluxes
are similar to those of other disks in spite of the low disk mass
bodes well for future studies. Targeted ALMA observations of
the strongest lines will allow much better sensitivity and are
expected to easily reduce the abundance limits by 1--2 orders of magnitude.
Together with searches for other complex organic molecules made
preferentially in the ice, this will allow direct tests of the
mechanism of sublimation of mid-plane ices in transitional disks
proposed by \citet{Cleeves2011}.

\section{Conclusions}

We observed the Oph IRS 48 protoplanetary disk with ALMA Early Science
at the highest frequencies, around 690 GHz, allowing the detection of
warm H$_2$CO and upper limits on the abundances of several other molecules including CH$_3$OH, H$^{13}$CO$^+$ , and CN
lines at unprecedented angular resolution.
\begin{enumerate}

\item We detected and spatially resolved the warm H$_2$CO
  9(1,8)-8(1,7) line, which reveals a semi-ring of emission at
  $\sim$60 AU radius centered south from the star. No emission is
  detected in the north. This demonstrates that H$_2$CO, an ingredient
  for building more complex organic molecules, is present in a
  location of the disk where planetesimals and comets are currently
  being formed.

\item The H$_2$CO emission was modeled using a physical disk model
  based on the dust continuum and CO emission \citep{BruderervdM},
  using three different trial abundance profiles. None of the profiles
  were able to match the observed data exactly, but the absolute flux
  indicates an abundance with respect to H$_2$ of $\sim10^{-8}$.

\item The combination of the H$_2$CO abundance in combination with
  upper limits for the CH$_3$OH emission indicates a H$_2$CO/CH$_3$OH
  ratio $>$0.3. This limit together with the overall abundance suggests
  that both solid-state and gas-phase processes occur in the disk.

\item Although the H$_2$CO emission is located only on the southern side of the disk, just like the millimeter
  dust continuum, the offset with the continuum peak and the low $S/N$ do not allow a firm claim on a relation with the dust-trapping
  mechanism.

\item The upper limit for H$^{13}$CO$^+$ indicates an HCO$^+$
  abundance of $<10^{-8}$, consistent with our model. The upper limit
  for CN of $10^{-7.3}$ relative to H$_2$ is directly at the level of
  that predicted by our model. Upper limits on the abundances of the other targeted molecules are consistent with earlier observations.

\item Future ALMA observations of intrinsically stronger lines will
  allow abundances to be measured that are one or more orders of magnitude below the
  upper limits derived here. This will allow full tests
  of the chemistry of simple and more complex molecules in
  transitional disks.
\end{enumerate}

  \begin{acknowledgements}
  The authors would like to thank C. Walsh for useful discussions and M. Schmalzl for help with the observational setup.
  N.M. is supported by the Netherlands Research School for Astronomy
  (NOVA), T.v.K. by the Dutch ALMA Regional Center Allegro financed by
  Netherlands Organization for Scientific Research (NWO) and
  S.B. acknowledges a stipend by the Max Planck
  Society. Astrochemistry in Leiden is supported by the Netherlands
  Research School for Astronomy (NOVA), by a Royal Netherlands Academy
  of Arts and Sciences (KNAW) professor prize, and by the European
  Union A-ERC grant 291141 CHEMPLAN. This paper makes use of the
  following ALMA data: ADS/JAO.ALMA\#2011.0.00635.S. ALMA is a
  partnership of ESO (representing its member states), NSF (USA) and
  NINS (Japan), together with NRC (Canada) and NSC and ASIAA (Taiwan),
  in cooperation with the Republic of Chile. The Joint ALMA
  Observatory is operated by ESO, AUI/NRAO and NAOJ.  
  \end{acknowledgements}

\bibliographystyle{aa}

\end{document}